\RequirePackage{silence}
\documentclass{aa}  
\usepackage{graphicx}
\usepackage{tabularx}
\usepackage{txfonts}
\usepackage{booktabs}     % For better table rulesv
\usepackage{array}        % For enhanced tables

\usepackage[version=4]{mhchem}
\usepackage{hyperref}
\usepackage{cleveref}
\usepackage{lastpage}
\usepackage{float}
\usepackage{longtable}
\usepackage{xtab}
\usepackage{placeins}
\usepackage{comment}
\usepackage{pdflscape}
\usepackage{siunitx}

\begin{document} 

    \title{
    A chemical perspective on planet formation in reduced systems}
    
   \author{Urja Zaveri\inst{1}
          \and Haiyang S. Wang\inst{2}
          \and Paolo A. Sossi\inst{1}
          }

    \institute{Institute of Geochemistry and Petrology, ETH Zurich, Switzerland\\
           \email{urja.zaveri@eaps.ethz.ch}, \email{paolo.sossi@eaps.ethz.ch}
         \and
           Centre for Star and Planet Formation, Globe Institute, University of Copenhagen, Denmark\\
           \email{haiyang.wang@sund.ku.dk}
         }

   \date{}

  \abstract
    {Relative abundances of refractory elements in planets are widely assumed to reflect those of their host stars. However, because elements are classified according to their behaviour in the solar nebula, this implicitly assumes that condensation is independent of nebular chemistry, despite evidence to the contrary in chemically reduced systems with high molar carbon-to-oxygen (C/O) ratios.}  
    {We investigate how variations in stellar C/O ratio and disk pressure modify condensation chemistry and assess the reliability of mapping stellar compositions to planetary building blocks in reduced environments.}
    {For a sample of FGK stars with C/O ratios spanning 0.65–0.95 (solar = 0.59$\pm$0.08), we compute equilibrium phase stability using \texttt{FactSage} over 1900–400 K at total pressures of $10^{-2}$, $10^{-4}$, and $10^{-6}$ bar. We track phase evolution and identify key chemical transitions across C/O, temperature, and pressure. Bulk planetesimal compositions are derived using a stochastic accretion framework that aggregates condensates from temperature-dependent feeding zones.}
    {We identify three distinct condensation regimes: (i) solar-like (C/O $\lesssim$ 0.7), (ii) transitional (C/O $\sim$ 0.7-0.91), and (iii) reduced (C/O $\gtrsim$ 0.92). Relative to solar sequences, oxygen-bearing silicates condense at lower temperatures in transitional- and reduced regimes, while carbides, silicides, and sulfides appear. Bulk planetesimal Fe/Mg, Fe/Si, and especially Fe/O ratios deviate substantially from their host stellar values in transitional- and reduced sequences, producing more diverse rocky building blocks within a single disk, ranging from metal-rich, C- and S-bearing bodies to more Earth-like compositions.}
    {Condensation sequences are not universal across stellar compositions. In reduced disks, elemental ratios commonly treated as refractory based on solar-system condensation temperatures may not reliably trace planetary bulk composition. The distinct building blocks produced in high-C/O systems therefore provide potential formation pathways for metal-enriched super-Mercury analogues and chemically distinct C- and S-rich rocky planets, expanding the diversity of terrestrial compositions beyond solar-system analogues.}
    

   \maketitle

\section{Introduction}\label{sec:intro}
Over the past decade, the focus of exoplanetary science has evolved from the mere detection and classification of exoplanets to detailed characterisation of their physical and chemical properties. Approximately 20\% of known exoplanets have both radial velocity and transit measurements available \citep{Hinkel2024HostExoplanets, Christiansen2025TheUsage}, enabling the derivation of both planetary radii and (minimum) masses. These parameters allow for the estimation of bulk densities and offer first-order insights into internal compositions. Mass-radius relationships for rocky planets assuming a tripartite model - comprising an iron core, silicate mantle, and a volatile envelope (liquid $\ce{H2O}$ + H-He gas) are commonly used to model interior structure \citep{Seager2007MassRadiusExoplanets, Dorn2015CanMeasurements, Zeng2016MASSRADIUSPREM}.

However, the modelling of internal structures and compositions of rocky exoplanets is known to be degenerate with mass-radius measurements alone \citep{Nellis2002PlanetaryConstraints, Stevenson2002IntroductionInteriors, Khan2008ConstrainingData, Dorn2015CanMeasurements}. A widely adopted approach to reduce degeneracy in inferring rocky exoplanet compositions is to assume that some key elemental ratios (e.g., Mg/Si and Fe/Mg) in the host star are identical to those of the planet, given that stars and planets form from the same molecular cloud \citep{Dorn2015CanMeasurements, Hinkel2018TheStars, Wang2019EnhancedExoplanets}. 
This approach has its foundations in the observation that the refractory, lithophile elements have abundances in the Earth's mantle that are in CI-chondritic ($\sim$solar) \textit{relative} proportions \citep{Ringwood1966ChemicalPlanets}. However, extending this practice to elements with higher volatilities is fraught with problems, since moderately volatile elemental abundances in the mantles of terrestrial planets are correlated with their 50\% nebular condensation temperatures ($T_c^{50\%}$), defined as the temperature at which half of the mass fraction of a given element condenses from the gas phase \citep{ONeill1998CompositionFormation, Lodders2003SolarElements}. These empirical observations imply that, relative to solar (CI-chondritic) composition, moderately volatile elements become increasingly impoverished in Earth's mantle with increasing volatility \citep{Palme2014CosmochemicalComposition, Wang2018ThePlanet, Wang2019TheAbundances, Braukmuller2019EarthsSource, Sossi2022StochasticEarth}. 

Such volatility trends are representative of some chondritic meteorites \citep{Braukmuller2018TheAlteration, Alexander2019QuantitativeChondrites}, rocky bodies such as the Moon \citep{Charnoz2021TidalVolatiles} and Mars \citep{Yoshizaki2020TheMars, Khan2022GeophysicalMars}, and are inferred for other planetary systems based on abundances in polluted white dwarf atmospheres \citep{Harrison2018PollutedMaterial}, supporting the notion that devolatilization is a common feature of rocky planet formation \citep{Wang2019TheAbundances, Calogero2025CanLoss}. As a result, a convenient assumption states that refractory elements ($T_c^{50\%} > 1400~\mathrm{K}$; Al, Ca, Ti) and major elements  ($1300 < T_c^{50\%} < 1400~\mathrm{K}$; Mg, Si, and Fe) are incorporated into planetary compositions in roughly stellar proportions, while the abundances of moderately volatile elements are expected to vary more substantially \citep{Kargel1993TheEarth, McDonough1995TheEarth, Palme2014SolarElements, Wang2018ThePlanet, Wang2019TheAbundances}. \citet{Spaargaren2023PlausibleNeighborhood, Guimond2024FromExoplanets} explored the chemical diversity of planetary compositions as a function of host star abundances using Sun-Earth devolatilization factors and showed that the Sun and Earth are close to the medians of bulk compositions of Sun-like (FGK main sequence) stars and postulated terrestrial-type exoplanets around them. 

However, in stellar systems with abundances that differ markedly from solar values, condensation sequences are not mandated to follow solar patterns. \citet{Timmermann2023RevisitingCompositions} systematically varied elemental abundances of Sun-like stars and, by performing equilibrium condensation calculations, showed that, at a given total pressure, the metallicity, Z (defined as all elements heavier than H and He), and the molar C/O ratio of the gas have the highest influence on $T_c^{50\%}$ and hence on the plausible compositions of the resulting planets. 
Both quantities vary significantly among sun-like stars: The metallicity, [Fe/H] (used as a proxy for Z), spans roughly between -1 and 1 dex \citep{Buder2019TheTGAS, Hinkel2014StellarCatalog}, and relative variations in elemental abundances-particularly C and O-produce stellar C/O ratios typically between 0.1 and 1 in the solar neighbourhood \citep{Pignatari2023TheStars, Brewer2016C/ONEIGHBORHOOD}. 

Previous studies have shown that in protoplanetary disks, carbon monoxide (\ce{CO_{(g)}}) is predicted to be the most thermodynamically stable oxygen-bearing gas molecule \citep{Larimer1975TheMaterial, Larimer1979TheChondrites}. The solar C/O ratio  of $\sim$0.59 $\pm$ 0.08 \citep{Lodders2003SolarElements, Asplund2021TheVision, Lodders2021RelativeGroups} results in small quantities of $\ce{CO_{(g)}}$, and the excess O forms $\ce{H2O_{(g)}}$ via equation:

\begin{equation}
    \ce{H2_{(g)}} + \ce{1/2O2_{(g)}} = \ce{H2O_{(g)}}
    \label{eq:H2_H2O}
\end{equation}

The resulting $f$O$_2$ is sufficient to allow the condensation of Al-, Ca-, Ti-, and Mg-bearing oxides and silicates, such as corundum, perovskite, and forsterite.

On the other hand, as C/O increases, oxygen becomes progressively tied up in $\ce{CO_{(g)}}$. With C/O approaching 0.8, the fugacities of other oxygen-bearing gas species (e.g., $\ce{SiO_{(g)}}$, $\ce{H2O_{(g)}}$) decline accordingly. When the partial pressure of $\ce{CO_{(g)}}$, $p$(CO), exceeds that required for graphite activity to reach unity, C-bearing species condense \citep{Larimer1975TheMaterial, Adams2025EffectsStars}. Such sequences lead to lower $f$O$_2$ than in Sun-like sequences and suppress the formation of oxide or silicate condensates in favour of reduced phases such as other sulfides, nitrides, and silicides. Consequently, the condensation sequence, and hence the bulk composition of the ensemble of condensed phases at any given $P-T$, diverges markedly from those produced from a solar composition \citep{Timmermann2023RevisitingCompositions, Spaargaren2025Proto-planetaryFormation}.

Determining how these compositionally distinct condensates affect bulk planet(esimal) compositions was undertaken by \citet{Bond2010THESIMULATIONS}, who combined equilibrium condensation calculations with N-body simulations, linking specific pressure-temperature (\(P\)–\(T\)) conditions in disks to planetesimal composition. Their simulations show that disks with C/O~\(>\)~0.8 yield carbon-rich solids, such as graphite, silicon carbide ($\ce{SiC}$), and titanium carbide ($\ce{TiC}$), dominating in the inner regions. In extreme cases (C/O $\sim$ 1–2), planet(esimals) may contain up to 75 wt\% carbon \citep{Bond2010THESIMULATIONS}. In a similar vein, \citet{Moriarty2014CHEMISTRYCOMPOSITION} introduced a sequential condensation model in which the composition of the residual bulk gas composition used in the subsequent (i.e., down-temperature) $P-T$ step changes upon cooling via the step-wise removal of condensed solids from the system. This model predicts that carbon-rich planet(esimals) can form even in initially oxygen-rich environments starting from C/O~\(\sim\)~0.65.

Building upon this framework, \citet{Shakespeare2025TheStars} implemented a dynamic, viscously evolving protoplanetary disk model coupled with sequential condensation. Holding all elemental abundances constant except for C and O, they identify three regimes of disk chemistry: silicate-dominated (C/O~\(\lesssim\)~0.6), transitional (C/O~\(\sim\)~0.6–0.9), and carbide-dominated (C/O~\(\gtrsim\)~0.9) based on the type and amount of condensates present. Their results show that small changes in C/O within the transitional regime lead to non-linear shifts in condensate identity, with reduced phases like graphite and $\ce{SiC}$ becoming dominant.

Furthermore, \citet{Spaargaren2025Proto-planetaryFormation} sampled about 1000 FGK dwarf stars and parametrised $T_c^{50\%}$ across a wide range of initial disk compositions, enabling disk-dependent devolatilization trends to be incorporated when estimating the bulk composition of rocky exoplanets. They find that Earth-like planets are expected to form in disks with C/O $\leq 0.75$, whereas more reduced planets emerge at C/O $> 0.75$.

Although existing models broadly reproduce consistent, increasingly carbide-, nitride- and silicide-rich condensation sequences to higher  C/O ratios, they do not focus on the underlying chemical pathways that control condensate stability in protoplanetary disks. Moreover, \cite{Spaargaren2025Proto-planetaryFormation} highlighted the infeasibility of determining a $T_c^{50\%}$ for O (and other similarly behaving elements) that condense partially over a wide range of temperatures, thereby undermining the use of $T_c^{50\%}$ as a robust proxy for predicting the compositions of rocky planets [See also \citet{Lodders2025CondensationEarth,Sossi2025PhysicochemicalPlanets}]. Consequently, a more fundamental approach to understanding the thermodynamic stability of these reactions is essential both for linking element availability to disk gradients and compositions and for developing future models that incorporate non-equilibrium processes such as kinetic activation energies for gas-solid reactions. Moreover, most models neglect \textit{(i)} solid solutions and \textit{(ii)} exotic condensates, including iron silicides and sinoite, which are absent under solar-like compositions but become critically important in reduced environments. Addressing these gaps requires a systematic investigation of high C/O systems within a framework that also enables direct tracking of how these variations map onto the bulk compositions of accreting planet(esimal)s, which cannot be inferred from condensation temperatures alone.

In this study, we aim to:
\begin{itemize}
    \item Identify the temperature- and pressure-dependent condensation reactions as a function of C/O ratio.
    \item Examine how the star-planet chemical link varies over the same range of C/O ratios.
    \item Quantify the impact of these condensates on the bulk composition and redox state of rocky planetary building blocks formed in chemically reduced environments.
\end{itemize}

%--------------------------------------------------------------------
\section{Data and methodology}\label{sec:data}

\subsection{Stellar chemistry}
We use a set of measured stellar abundances with varied C/O values as proxies for initial disk compositions. This approach preserves nucleosynthetic correlations between elements, in contrast to arbitrarily modifying solar values.
The stellar abundance data of Sun-like stars (FGK dwarfs) measured on the HARPS GTO spectrograph and compiled from \cite{Adibekyan2012ChemicalProgram} and subsequent works \citep{DelgadoMena2017ChemicalProgram, CostaSilva2020ChemicalSample, DelgadoMena2021ChemicalProgram}, are collectively referred to as the Portuguese Group Dataset (PGD) here. The authors derived the abundances in this dataset using a local thermodynamic equilibrium analysis, adopting the Sun as the reference and employing the spectral synthesis code MOOG together with a grid of Kurucz ATLAS9 model atmospheres \citep{Adibekyan2012ChemicalProgram}. For further details, we refer the reader to the references cited above. The selection of this dataset is further rationalised in \autoref{sec:stellar abundance}.

We then exclude stars with $\log g \leq 3.5$ to avoid giant stars and filter out stars with effective temperatures $T_{\text{eff}} < 5000$K due to systematic trends in certain elements (S, Si, Na, and Al) at lower temperatures [see \citet{Adibekyan2012ChemicalProgram} and \autoref{sec:stellar abundance}]. We further exclude stars without abundance measurements for any of the elements within the Ni-Fe-Cr-Ti-Ca-S-Si-Al-Mg-Na-O-C system, leading to a sample of 532 stars.

\begin{figure}[!ht]
    \centering
    \includegraphics[width=1\linewidth]{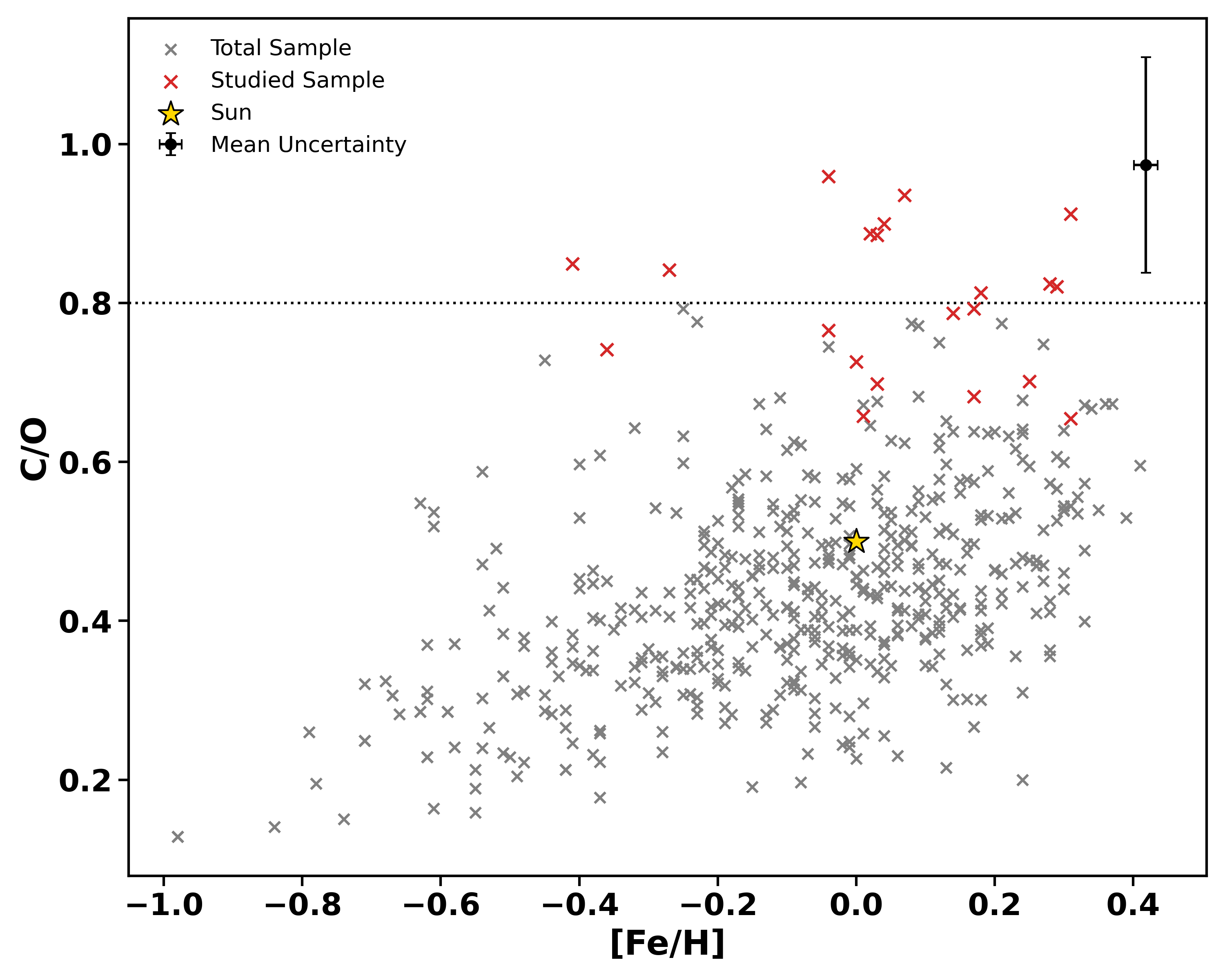}
    \caption{Distribution of our adopted dataset of elemental abundances of Sun-like stars (FGK dwarfs) on the [Fe/H]-C/O diagram. Sample of stars selected for this study is highlighted in red. The solar reference \citep{Lodders2003SolarElements} is shown as a star sign. The typical error bars of the dataset in the [Fe/H]-C/O space are shown in the upper right corner.}
    \label{fig:PGD_data_points}
\end{figure}

From this subset, we include all available stars with C/O $> 0.8$ since this is the threshold recognised as marking a major shift in disk chemistry \citep{Larimer1975TheMaterial, Bond2010THESIMULATIONS, Shakespeare2025TheStars}. In addition to these, we also sampled several stellar compositions in C/O range in the range of C/O$=0.65$–$0.8$ at varying metallicity to explore the behaviour of disks approaching the reduced regime. The full dataset and the selected samples used in this study are shown in \autoref{fig:PGD_data_points}.

The PGD lacks nitrogen abundances. Nitrogen is important for such modelling because, although it does not condense below $400$~K in solar-like disks, it is expected to condense into nitrides in high C/O systems. We therefore estimate N by scaling it relative to O based on the argument made by \citet{Nicholls2017AbundanceGalaxies} who show that N does not scale linearly with metallicity due to its mixed origins from primary core-collapse and secondary dredge-up processes in stars. We use two independent datasets containing nitrogen and oxygen abundances (\citet{Magrini2018TheWay} and \citet{DaSilva2024ArielStars}) to derive second-order polynomial relationships between absolute abundances A(O) and A(N) using orthogonal distance regression (ODR), where absolute abundances are defined on the standard astronomical logarithmic scale, $A(X) = \log_{10}(N_X/N_H) + 12$.
Applying the polynomial fit to the selected dataset for this study, the two fits differ by a mean offset (bias) of +0.03 dex and a root-mean-square-error of 0.18 dex. We use the fit obtained from \citet{Magrini2018TheWay} dataset as shown in \autoref{eq:ODR-N} to estimate absolute nitrogen abundances because it reproduces the solar nitrogen abundances from \citet{Lodders2003SolarElements} with better consistency than the other dataset. 
\begin{equation}
A(N) = -1.6665 \cdot A(O)^2 + 31.2043 \cdot A(O) - 137.4157
\label{eq:ODR-N}
\end{equation}
The resulting nitrogen abundances are subsequently converted to elemental number fractions (mol\%) and incorporated into the dataset along with the remaining elements. The final composition of the stars considered in this study is provided in \autoref{tab:Stellar_data}. 

%%%%%%%%%%
\subsection{Gibbs free energy minimisation}
To calculate the stability and abundances of gas species and condensed phases (liquid, solid) in thermodynamic equilibrium, we use the \texttt{Equilib} module of \texttt{FactSage 8.2} \citep{Bale2016FactSage2010-2016}. The calculations for a given composition are performed at constant total pressure, over a temperature range of 1900~K to 400~K. To accurately capture condensation and gas-mineral reactions, we use a small temperature step of $5$~K. We determine the equilibrium phase assemblage for each temperature step by Gibbs energy minimisation. We incorporate often-neglected solid species such as $\ce{Fe3C}$, $\ce{Fe_xSi}$ $(x = 1, 3)$, $\ce{Al4C3}$, and $\ce{Si2N2O}$, which do not condense in the solar nebula but may condense in reduced systems. In total, the calculations involve a selection of 199 gases, 50 solids, and 38 solid solutions, as provided in \autoref{sec:Factsage_species}.

\subsection{Synthetic planet(esimal) model} \label{sec:accretion_method}
\citet{Sossi2022StochasticEarth} demonstrate that, for Earth and Vesta, the abundance of an element $i$ correlates with its $T_c^{50\%}$ in a manner that can be approximated by a logistic-shaped cumulative distribution function. For example, Earth and Vesta differ in the mean temperatures experienced by their building blocks (parametrised by the inflexion point of the logistic curve, $T_0$) and in the range of temperatures from which material was collected (characterised by the steepness of the logistic function, $\sigma$).

Building on this empirical model, we estimate the bulk compositions of synthetic planet(esimals) by integrating the equilibrium condensates derived from our nebular condensation calculations over a range of temperatures. This could be related to a temperature gradient across the disk, or phases condensed over a range of times at a given location in a cooling disk.

For each element $i$, we calculate the mole fraction by integrating the product of the condensed material fraction and the normalised probability density function at each temperature point within the range of 400~K to 1900~K, centred around a specified $T_0$ with a specified $\sigma$.  \newline

For a given $T_0$ and $\sigma$, the probability density function of a normal distribution, $f(T)$, over the temperature range $T$ is:
\begin{equation}
f(T) = \frac{1}{\sigma \sqrt{2\pi}} \exp\left( -\frac{(T - T_0)^2}{2\sigma^2} \right),
\label{eq:PDF}
\end{equation}

The normalized PDF, $\phi(T)$, is given by:
\begin{equation}
\phi(T) = \frac{f(T)}{\sum_{T_c} f(T)},
\label{eq:normalized_PDF}
\end{equation}
where, the summation ($\sum_{T_c} f(T)$) is performed over the temperature range, $T_c$, defined by a given $T_0$ and $\sigma$ that represents a Gaussian-like feeding zone.

The molar planetary abundances of an element $i$, denoted by $X_i$, are calculated as:
\begin{equation}
X_i = \sum_T \eta_i(T) \cdot \phi(T),
\label{eq:planetary_abundances}
\end{equation}

Where $\eta_i(T)$ represents the condensed molar fraction of element $i$ at temperature $T$, and $\phi(T)$ is the normalised PDF. \newline

\citet{Sossi2022StochasticEarth} report a mean temperature for the Earth of $T_0 = 1144$~K with $\sigma = 225$~K. In our study, we study the effect of varying $T_{0}$ across 650, 900, 1150, and 1400~K while keeping $\sigma = 225 \text{K}$ fixed to model the accretion of Earth-mass planetary bodies around these exotic stars. The effect of varying $\sigma$ on planetary composition is also examined, confirming that our principal conclusions are not contingent on this choice.
%%%%%%%%%%%%%%%%%%%%%%%%%%%%%%%%%%%%%%%%%%%%%%%%%%%%%%%%%%%%

\section{Results}\label{sec:Results}
\subsection{General behaviour and classification}
The equilibrium condensation sequences calculated at a disk pressure of $10^{-4}$ bar for our selected stellar compositions (with C/O ranging from 0.6 to 0.95), reveal three distinct regimes of condensate mineralogy:

\textbf{(i) Solar-like sequence (C/O $\leq 0.7$)} - Condensation proceeds through reactions analogous to those in the solar nebula. The assemblage consists exclusively of oxides and silicates.

\textbf{(ii) Transitional sequence (C/O $\sim 0.7 - 0.91$)} - Reducing phases, primarily graphite but also iron silicides, sulfides, carbides, and nitrides, begin to appear alongside the oxides and silicates present in the solar-like sequences. Their presence leads to the onset of silicate and oxide condensation at lower temperatures.

\textbf{(iii) Reduced sequence (C/O $\geq 0.92$}) - characterised by little- to no silicate or oxide formation in the refractory temperature range ($T > 1400$~K at $10^{-4}$~bar), and defined by the first appearance of $\ce{SiC_{(s)}}$ as a direct condensate (here at $\sim$1510~K). The other, highest temperature refractory condensates are $\ce{TiC_{(s)}}$ or graphite. The specific reactions responsible for this behaviour are discussed in section \ref{sec:carbon}.

These results indicate that the transitions between solar-like, transitional and reduced sequences occur near C/O ratios of $\sim$0.7 and $\sim$0.92, respectively. While the C/O ratio provides the primary classification metric, we note that, for a fixed C/O, higher metallicity generally promotes the earlier condensation of reducing phases. This is because additional metals consume oxygen, leaving less O available to form silicates and oxides \citep{Larimer1967ChemicalElements, Adams2025EffectsStars}.

As archetypes of condensation sequences in non-solar compositions, we highlight HD94151 (C/O = 0.89) for the transitional sequences and HD24633 (C/O = 0.95) for the reduced sequences. We note that not all transitional sequences necessarily condense every reduced phase in addition to graphite, as this depends on both metallicity and C/O ratio. For illustration of reactions, we use the sequence HD94151 (C/O =0.89) as an example, since its relatively high C/O ratio and metallicity lead to the condensation of all phases relevant to transitional sequences. HD24633 was used for the reduced sequence, due to it being the highest C/O ratio-bearing star in our dataset. Both sequences are compared against our model of condensation of the solar nebula using the composition of \citet{Lodders2003SolarElements}, as described in section \ref{sec:disk_chem_composition}, to demonstrate the evolving nature of disk chemistry, and hence, condensation sequence, across these regimes.

In section \ref{sec:disk_chem_pressure}, we explore the variation in composition due to changes in total pressure, constant as a function of $T$ at $10^{-6}$ bar, $10^{-4}$ bar, or $10^{-2}$ bar. Finally, the resultant bulk planetary compositions arising from variations in both composition and pressure are presented in section \ref{sec:planet_composition}.

\subsection{Condensate mineralogies with varying C/O ratios} \label{sec:disk_chem_composition}

\begin{figure}[!ht]
    \centering
    \includegraphics[width=1\linewidth]{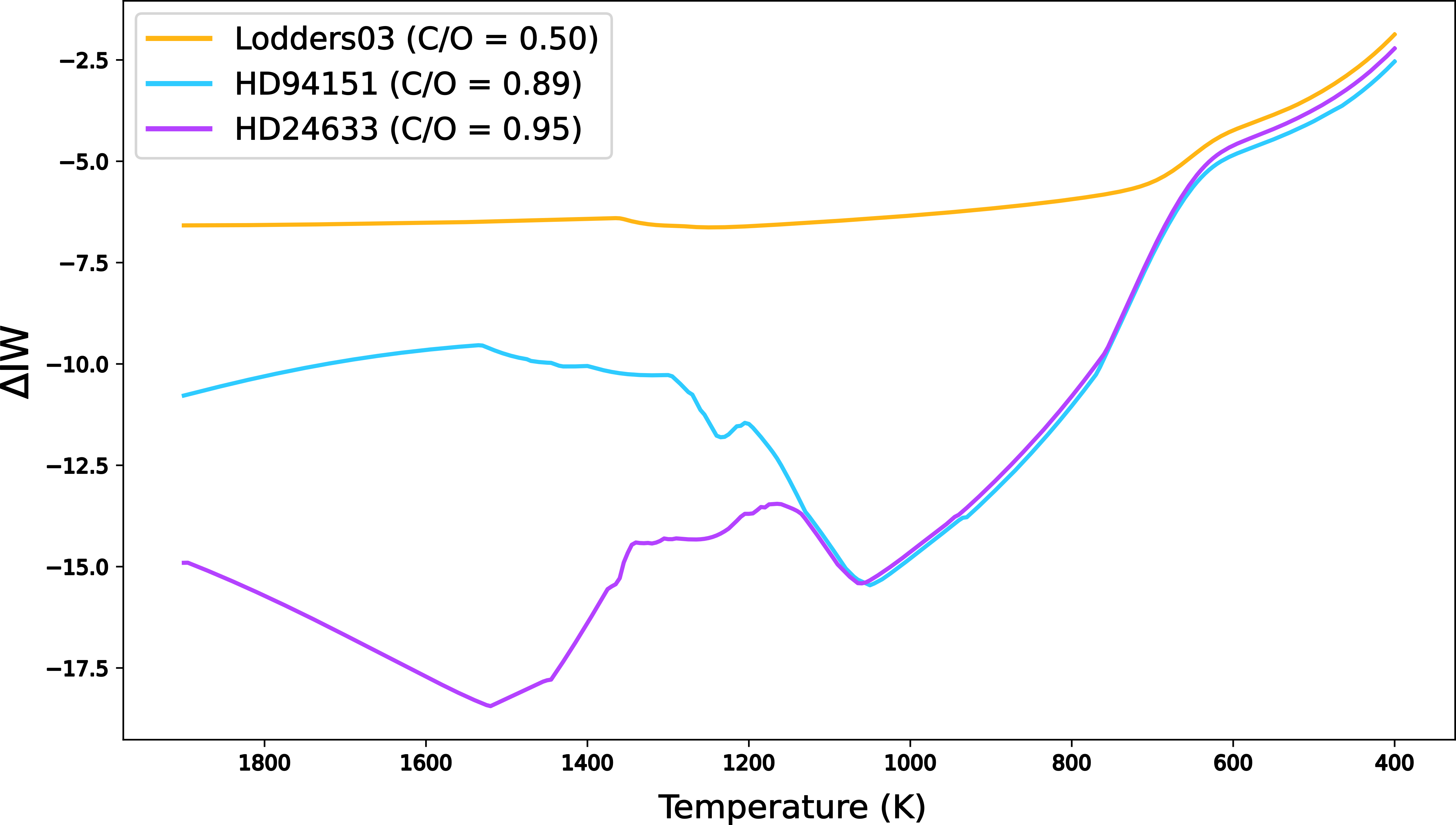}
    \caption{The variation of oxygen fugacity with respect to the Iron-W\"ustite buffer ($\Delta$IW) as a function of disk mid-plane temperature for three representative cases: Lodders03 (`solar-like sequence'), HD94151 (`transitional sequence'), and HD24633 systems (`reduced sequence').}
    \label{fig:delta_iw_plot_final}
\end{figure}

The chemical differences due to system's C/O ratio are reflected in its oxygen fugacity ($f$O$_2$) at a given pressure and temperature. Oxygen fugacity is primarily set by the $\ce{H2O_{(g)}}/\ce{H2_{(g)}}$ ratio ( \autoref{eq:H2_H2O}) and hence is higher at lower C/O. 

\autoref{fig:delta_iw_plot_final} illustrates the variation in oxygen fugacity ($f$O$_2$) relative to the Iron--Wüstite buffer ($\Delta$IW) for three representative cases: Lodders03 (C/O = 0.50; `solar-like sequence'), HD94151 (C/O = 0.89; `transitional sequence'), and HD24633 (C/O = 0.95; `reduced sequence'). The $f$O$_2$ of Iron--Wüstite is evaluated as a function of temperature using the standard thermodynamic parametrisation for the Fe-FeO equilibrium from \citet{ONeill1993ThermodynamicBuffer} (their Eq. 8).
At higher temperatures, HD94151 and HD24633 display lower $f$O$_2$ compared to those calculated for a solar gas at the same pressure and temperature. The troughs in the graph correspond to points at which the $f$\ce{H2O}/$f$\ce{H2} ratio decreases due to condensation reactions detailed in the subsequent subsections. Because these reactions differ among the three canonical sequences, the $\Delta$IW values span a range of 10 orders of magnitude at high temperatures ($>$ 1400~K). By $\sim$1100~K, the $\Delta$IW values of the transitional- and reduced sequences become indistinguishable from one another. Notably, at temperatures below $700~\mathrm{K}$, the $\Delta$IW values of all three systems converge similarly, suggesting comparable redox conditions independent of C/O ratio below this temperature.

\begin{figure*}[!ht]
    \centering
    \includegraphics[width=0.8\linewidth]{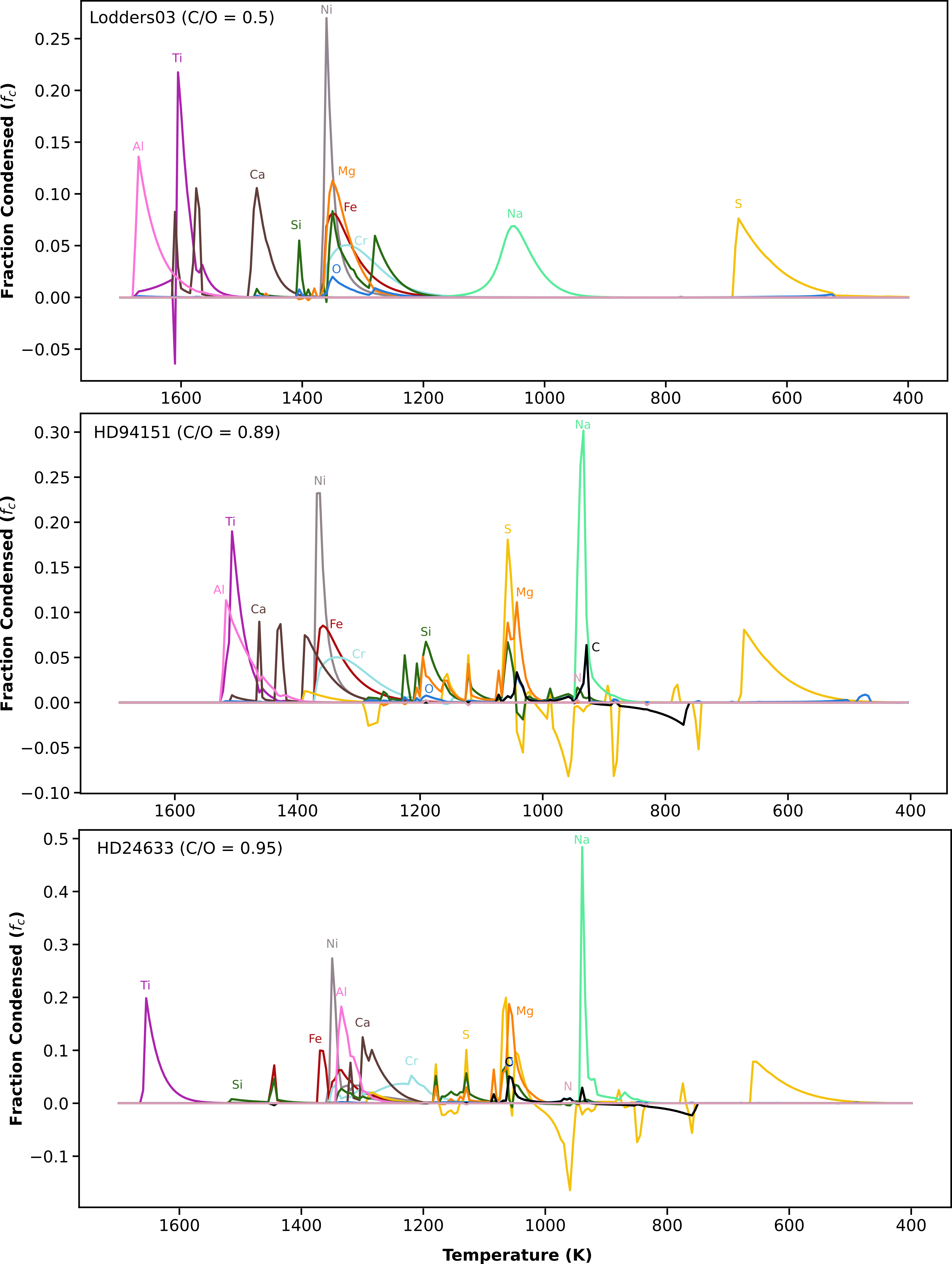}
    \caption{Instantaneous fraction condensed ($f_c$) of each element as a function of temperature in solar-like (Lodders03; C/O$= 0.50$), transitional (HD94151; C/O$= 0.89$), and reduced (HD24633; C/O$= 0.95$) systems at $10^{-4}$ bar total pressure. For relevant reactions corresponding to the condensation peaks, see subsections \ref{sec:Refractory}--\ref{sec:sulfide}}
    \label{fig:fc_curves_updated_final}
\end{figure*}

\autoref{fig:fc_curves_updated_final} shows the instantaneous condensed fraction of each element j, ($f_c^j$), in a monotonically cooling system ($T_{i-1} > T_i$), calculated as:

\begin{equation} \label{condensed_fraction}
    f_c^{j} = x^{j}_{solid} (T_{i}) - x^{j}_{solid}(T_{i-1})
\end{equation}

Here, $x^j_{solid}(T)$ is the cumulative molar fraction of element j present in condensed phases at temperature $T$.
The positive peaks represent the initial condensation of major phases directly from the gas phase, while the troughs indicate the re-evaporation of a condensed solid. It is important to note that these curves do not account for the transformation of one solid to another through subsequent solid-state reactions, but rather only the fraction of the element condensed from the gas phase. 

\autoref{tab:fc-temperatures} lists the peak temperatures (i.e $
\frac{d f_c^2}{d T^2} = 0
$) associated with the initial condensation of distinct condensate phases in the solar-like (Lodders03; C/O$= 0.50$), transitional (HD94151; C/O$= 0.89$), and reduced (HD24633; C/O$= 0.95$) systems, calculated at the canonical pressure of $10^{-4}$ bar. Only the first major condensation peaks are included; secondary peaks corresponding to the re-condensation of previously condensed phases are not reported here. Peaks with $f_c < 0.04$ have been omitted, as they do not represent significant contributions- except in the case of nitrogen, which is retained to emphasise the partial condensation of N in high C/O systems.

In solar-like systems, most elements exhibit a single, well-defined condensation peak corresponding to most of their incorporation into solid(s). Exceptions to this generality include \ce{Ca}, \ce{Si}, and \ce{O}, which display more complex condensation behaviour. Notably, \ce{N} and \ce{C} do not condense above $400~\mathrm{K}$ in the solar nebula, while \ce{S} condenses only below $700~\mathrm{K}$, predominantly in the form of troilite (\ce{FeS_{(s)}}). In transitional and reduced systems, \ce{N} (as sinoite and TiN) and \ce{C} (as SiC and C) condense and \ce{S} condenses not only into troilite, but also at higher temperatures via oldhamite and niningerite (\ce{(Ca,Mg)S_{(s)}}). Sulfur peaks show complex behaviour due to multi-stage condensation–evaporation cycles.

\begin{table*}[!htbp]
\centering
\caption{Peak fractional crystallisation ($f_c$) and condensation temperatures ($T$) for solar-like (Lodders03; C/O$= 0.50$), transitional (HD94151; C/O$= 0.89$), and reduced (HD24633; C/O$= 0.95$) systems, calculated at disk pressure of $10^{-4}$ bar.}\vspace{0.5em}
\label{tab:fc-temperatures}
\begin{tabular}{llllllll}
\hline
\textbf{Major Phase} & \textbf{Element} & \multicolumn{2}{c}{\textbf{Lodders03 (solar-like)}} & \multicolumn{2}{c}{\textbf{HD94151 (Transitional)}} & \multicolumn{2}{c}{\textbf{HD24633 (Reduced)}} \\ 
\cmidrule(lr){3-4} \cmidrule(lr){5-6} \cmidrule(lr){7-8}
                    &                   & $f_c$ (mol \%) & $T$ (K)  & $f_c$ (mol \%) & $T$ (K)  & $f_c$ (mol \%) & $T$ (K) \\ \hline
Corundum*           & Al               & 0.136 & 1670 & 0.113 & 1525 & 0.183 & 1335 \\
Perovskite*         & Ti               & 0.105 & 1575 & 0.190 & 1515 & - & - \\
TiC                 & Ti               & - & - & - & - & 0.198 & 1655 \\
CaS*                & Ca               & - & - & 0.075 & 1395 & 0.080 & 1290 \\
Grossite*           & Ca               & 0.105 & 1575 & 0.087 & 1435 & 0.125 & 1300 \\
Hibonite*           & Ca               & 0.083 & 1610 & 0.090 & 1470 & 0.077 & 1320 \\
Ca-oxide slag*      & Ca               & 0.106 & 1475 & - & - & - & - \\
Fe-$\textit{fcc}$*             & Fe               & 0.082 & 1350 & 0.085 & 1365 & 0.063 & 1335 \\
Fe$_3$Si            & Fe               & - & - & 0.053 & 1230 & 0.100 & 1370 \\
FeSi                & Fe               & - & - & - & - & 0.072 & 1445 \\
Fe-$\textit{fcc}$*             & Ni               & 0.270 & 1360 & 0.232 & 1370 & 0.274 & 1350 \\
\ce{Cr4C}           & Cr               & - & - & - & - & 0.052 & 1220 \\
Fe-$\textit{fcc}$*             & Cr               & 0.050 & 1325 & 0.050 & 1340 & - & - \\
Clinopyroxene*      & Mg               & - & - & 0.0508 & 1200 & - & - \\
Olivine*            & Mg               & 0.113 & 1350 & - & - & 0.064 & 1085 \\
Ca-Feldspar*        & Si               & 0.055 & 1405 & - & - & - & - \\
Clinopyroxene*      & Si               & - & - & 0.043 & 1210 & - & - \\
Fe$_3$Si            & Si               & - & - & 0.052 & 1230 & - & - \\
FeSi                & Si               & - & - & - & - & 0.046 & 1445 \\
Olivine*            & Si               & 0.083 & 1350 & 0.067 & 1195 & 0.057 & 1130 \\
Orthopyroxene*      & Si               & 0.059 & 1280 & - & - & - & - \\
Orthopyroxene*      & O                & - & - & 0.059 & 930 & - & - \\
Ti-spinel*          & O                & - & - & - & - & 0.049 & 1060 \\
Na-Feldspar*        & Na               & 0.069 & 1050 & 0.302 & 935 & 0.484 & 940 \\
TiN                 & N                & - & - & 0.003 & 1195 & 0.001 & 1195 \\
MgS*                & S                & - & - & 0.053 & 1125 & 0.074 & 1180 \\
Troilite*           & S                & 0.076 & 680 & 0.081 & 670 & 0.079 & 655 \\
Graphite            & C                & - & - & 0.0637 & 930 & 0.051 & 1060 \\ \hline
\end{tabular}
\tablefoot{Phases marked with an asterisk (*) were treated as solid solutions in the thermodynamic calculations. Only phases that directly condense from the gas phase exhibit $f_c$ peaks. Additionally, the table highlights only the first peak associated with the condensation of significant phases - namely, any recondensation of such phases is not reported here.}
\end{table*}

In the following subsections, we discuss condensation sequence and disk chemistry that reflect these elemental condensation peaks, as well as variations in oxygen fugacity across systems with varying C/O.  

%%%%%%%%%%%%%%%%%%%%%%%%%%%%%%%%%%%%%%%%%%%%%%%%%%%%%%%%%%%%%%%

\subsubsection{Oxide phases}\label{sec:Refractory}

\begin{figure*}[ht!]
    \centering
    \includegraphics[width=1\linewidth]{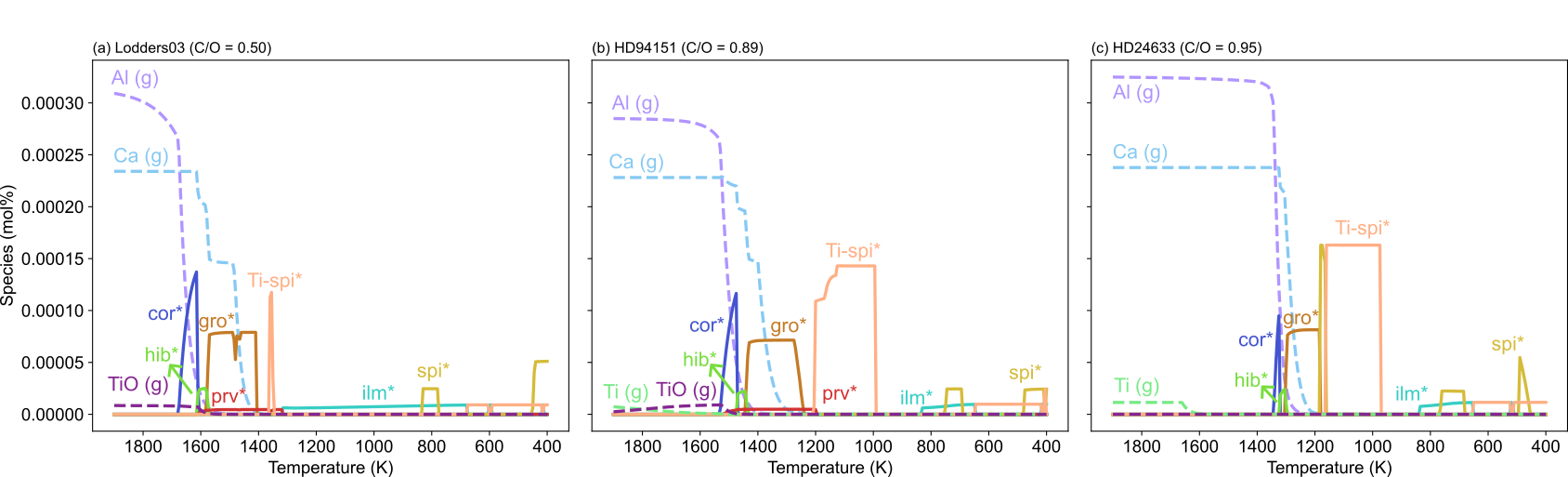}
    \caption{Condensation of (refractory) oxides (Al, Ca, Ti, O $\pm$ Mg, Fe, Cr, Ni) in solar-like (Lodders03; C/O$= 0.50$), transitional (HD94151; C/O$= 0.89$), and reduced (HD24633; C/O$= 0.95$) systems at a disk pressure of $10^{-4}$ bar. Species marked with an asterisk (*) are treated as solid solution phases in the calculations.}
    \label{fig:refractory_comp_linear}
\end{figure*}

In solar-like nebular environments, refractory elements (here Ca, Al and Ti) mostly condense as oxide phases. 
As shown in \autoref{fig:refractory_comp_linear}(a), corundum (cor*) - \ce{Al2O3} - initiates the condensation sequence (\ce{Al} $f_c$ peak: $1670~K$) via the net chemical reaction:
\begin{equation}
\ce{2Al_{(g)} + 3H2O_{(g)} -> Al2O3_{(s)} + 3H2_{(g)}}
\label{eq:corundum}
\end{equation}
or equivalently
\begin{equation}
\ce{2Al_{(g)} + 3/2 O2_{(g)} -> Al2O3_{(s)}}
\label{eq:corundum_alt}
\end{equation}

As the gas cools, \ce{Ca_{(g)}} reacts with condensed corundum to produce Ca-rich hibonite (hib*) - \ce{CaAl12O19} (\ce{Ca} $f_c$ peak: 1610~K) via net chemical reaction:
\begin{equation}
    6\ce{Al2O3_{(s)}} + \ce{Ca_{(g)}} + \frac{1}{2}\ce{O2_{(g)}} = \ce{CaAl12O19_{(s)}}
\label{eq:hibonite}
\end{equation}

Hibonite is then consumed to form Ca-rich grossite (gro*) - \ce{CaAl4O10} (\ce{Ca} $f_c$ peak: $1575~K$) via net chemical reaction:
\begin{equation}
 \ce{CaAl12O19_{(s)}} + 2\ce{Ca_{(g)}} + \ce{O2_{(g)}} = 3\ce{CaAl4O7_{(s)}}
\label{grossite} 
\end{equation}

Titanium, present primarily as $\ce{TiO_{(g)}}$, first condenses as $\ce{CaTiO3_{(s)}}$ - Ca-perovskite (prv*)  with \ce{Ti} $f_c$ peak: $1575~K$, through the net chemical reaction:

\begin{equation}
\ce{TiO_{(g)} + Ca_{(g)} + O2_{(g)} -> CaTiO3_{(s)}}
\label{eq:perovskite}
\end{equation}

Below $\sim1400$~K, further condensation proceeds through heterogeneous gas–solid reactions, leading to the formation of additional oxide phases including \ce{(Mg,Al,Ti,Fe)3O4_{(s)}} - Ti-spinel (Ti-spl*), \ce{(Fe,Mg,Al,Ti)(Fe,Al,Ti)O3_{(s)}} - ilmenite (ilm*) and \ce{(Fe,Mg,Ni,Cr,Al)(Fe,Al,Cr,Ni,Mg)2O4} Cr-bearing spinel (spi*)

Transitional sequences (\autoref{fig:refractory_comp_linear}(b)) exhibit a comparable phase assemblage to solar-like sequences but shifted to lower temperatures. Condensation of corundum (Al $f_c$ peak: 1525~K) is followed by perovskite (Ti $f_c$ peak: 1515~K), hibonite (Ca $f_c$ peak: 1470~K) and grossite (Ca $f_c$ peak: 1435~K). $\ce{TiO_{(g)}}$ remains the dominant titanium-bearing gas at- and below the temperature of the appearance of the first condensate but \ce{Ti_{(g)}} predominates at higher temperatures.  Ti-spinel is stable over a wider temperature range here (1200–1000 K) than in solar-like cases. In solar-like systems, these elements normally combine with Si to make silicates. Here, however, Si instead forms iron silicides (section \ref{sec:metal}) due to the lower $f$O$_2$, meaning less Si is incorporated into silicates. Below 900~K, phases Ti-spinel, ilmenite, and spinel form in a manner reminiscent of the solar-like sequences.

\autoref{fig:refractory_comp_linear}(c) highlights the \textbf{reduced sequences}, where no oxides and silicates condense in the temperature range of refractory phases in the solar sequences ($> 1400$~K). As $\ce{Ti_{(g)}}$ overtakes $\ce{TiO_{(g)}}$ as the primary titanium-bearing gas, Ti condenses as a carbide phase instead of Ca-perovskite due to low $\ce{TiO_{(g)}}$ partial pressure [for more details see section \ref{sec:carbon}]. Ca- and Al- bearing oxides condense similarly to the other two sequences but at still lower temperatures: corundum condensation (\ce{Al} $f_c$ peak: 1335~K) is followed by hibonite (\ce{Ca} $f_c$ peak: 1320~K) and grossite (\ce{Ca} $f_c$ peak: 1300~K). Similar to transitional sequences, Ti-spinel is stable over a wide temperature range, whereas oxides forming below 900~K mirror those in other sequences (again due to converging $f$O$_2$-$T$ paths, cf. \autoref{fig:delta_iw_plot_final}).

\subsubsection{Major silicates}\label{sec:silicate}
\begin{figure*}[ht!]
    \centering
    \includegraphics[width=1\linewidth]{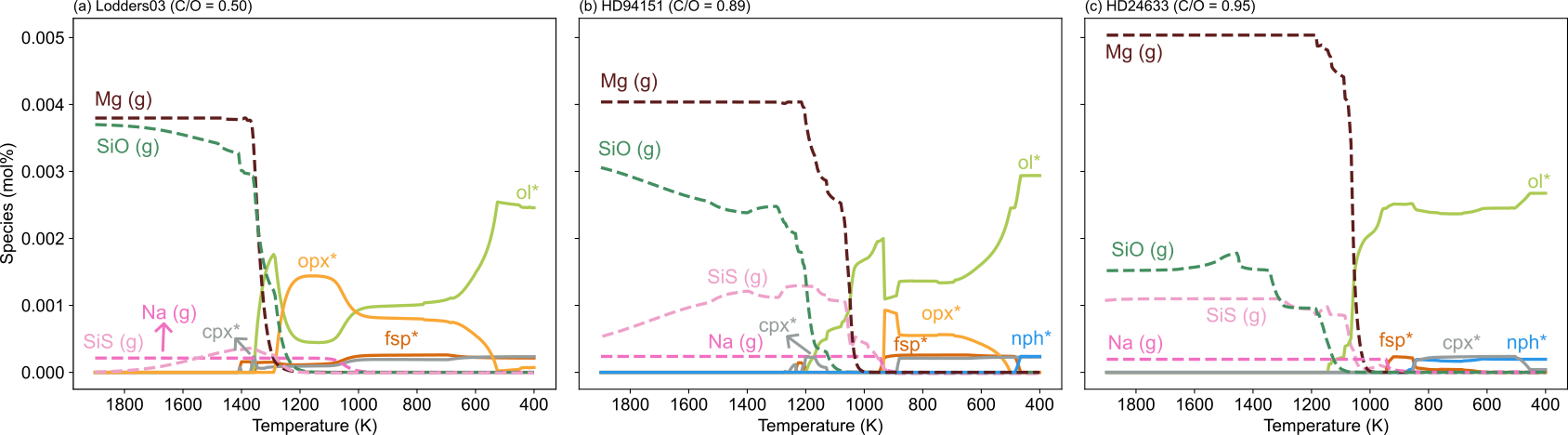}
    \caption{Same as \autoref{fig:refractory_comp_linear}, but for condensation of major silicate phases.}
    \label{fig:silicate_comp_linear}
\end{figure*}

While numerous minor silicate species condense, we focus here on the major silicates that remain stable over a temperature range of at least 200~K.
The major silicate bearing gas in sun-like sequences is $\ce{SiO_{(g)}}$ followed by $\ce{SiS_{(g)}}$. 
\autoref{fig:silicate_comp_linear}(a), illustrates that silicate condensation in solar-like sequences begins with \ce{Ca(Mg,Fe,Al)Si2O6_{(s)}} - Ca-rich clinopyroxene (cpx*) and \ce{CaAl2Si2O8_{(s)}} - anorthitic feldspar (fsp*) - (\ce{Si} $f_c$ peak: 1405~K). The dominant solid solution components of silicate minerals are shown in \autoref{fig:silicate_solidsoln_comp_linear}.
Next, \ce{Mg2SiO4_{(s)}} - forsteritic olivine (ol*) condenses (\ce{Si} $f_c$ peak: 1350~K) through the net chemical reaction:

\begin{equation}
\ce{2Mg_{(g)} + SiO_{(g)} + 3H2O_{(g)} -> Mg2SiO4_{(s)} + 3H2_{(g)}}.
\label{eq:olivine_condensation}
\end{equation}

Subsequently, $\ce{MgSiO3_{(s)}}$ - enstatitic orthopyroxene (opx*) forms (\ce{Si} $f_c$ peak: 1280~K) at the expense of olivine via net chemical reaction:

\begin{equation}
\ce{Mg2SiO4_{(s)} + SiO_{(g)} + H2O_{(g)} -> 2MgSiO3_{(s)} + H2_{(g)}}.
\label{eq:orthopyroxene_condensation}
\end{equation}

As shown in \autoref{fig:silicate_solidsoln_comp_linear}, the feldspar (fsp*) becomes more albitic ($\ce{NaAlSi3O8}$) at cooler temperatures (\ce{Na} $f_c$ peak: 1050~K) . 

Similarly, as established in \citet{Grossman1972CondensationNebula, Mokhtari2026CondensationEnvironments} the fayalite component of olivine ($\ce{Fe2SiO4_{(s)}}$) increases below $\sim$600~K via the net chemical reaction:

\begin{multline}
    \ce{2Fe_{(s)} + 2MgSiO3_{(s)} + 2H2O_{(g)} ->} \\
    \ce{Mg2SiO4_{(s)} + Fe2SiO4_{(s)} + 2H2_{(g)}}
\label{eq:fayalite_reaction}
\end{multline}

\noindent\autoref{fig:silicate_comp_linear}(b) shows that in transitional sequences, the most abundant gas is still $\ce{SiO_{(g)}}$, followed by an increased stability of $\ce{SiS_{(g)}}$ at higher temperatures. Condensation begins with clinopyroxene ($\ce{(Ca,Al,Mg)Si2O6_{(s)}}$; Mg $f_c$ peak: 1200~K) and anorthite ($\ce{CaAl2Si2O8_{(s)}}$). Compared to solar-like sequences, anorthite stability decreases in both the abundance and the temperature of its first appearance \autoref{fig:silicate_solidsoln_comp_linear}(a, b). This is because some of the Ca is now sequestered in \ce{(Ca,Mg)}S due to the lower $f_{\ce{O2}}$ [See section \ref{sec:sulfide}]. 

The next silicate phases to condense are forsteritic olivine (\ce{Si} $f_c$ peak: 1195~K) and enstatitic orthopyroxene (\ce{O}$ f_c$ peak: 930~K) via \autoref{eq:olivine_condensation} and \autoref{eq:orthopyroxene_condensation}. The plagioclase becomes more albitic (\ce{Na} $f_c$ peak: 935~K). Enstatite breaks down to olivine + clinopyroxene  at $\sim$900 $K$. Fayalitic olivine becomes stable below $\sim$600~K, similar to solar-like sequences (\autoref{fig:silicate_solidsoln_comp_linear}). Notably, $\ce{NaAlSiO4_{(s)}}$ - nepheline (nph*) replaces feldspar below 500~K \autoref{fig:silicate_solidsoln_comp_linear}. 
The $f_c$ peaks of the condensed silicates in transitional sequences are lower compared to those in solar-like sequences.

As observed in \autoref{fig:silicate_comp_linear}(c), reduced sequences exhibit a further decline in $\ce{SiO_{(g)}}$ stability, with $\ce{SiS_{(g)}}$ being more abundant at higher temperatures, than was the case for transitional sequences. The first major silicate to condense is olivine (\ce{Si} $f_c$ peak: 1130~K), followed by limited condensation of plagioclase feldspar that contains both Ca and Na  (\ce{Na} $f_c$ peak: 940~K). The olivine solid solution becomes more fayalitic only below 500~K. Clinopyroxene and nepheline form via the breakdown of olivine and feldspar below $\sim$850K. Orthopyroxene does not condense in the system HD24633 due to its high Mg/Si ratio, but is observed in the other reduced system - HD68607. This behaviour is consistent with that reported by \citet{Jorge2022FormingComposition} for Star 3 (HIP 63048), where orthopyroxene is likewise absent at a comparable Mg/Si ratio.

It is important to note that the presence of nepheline in the systems HD94151 and HD24633 is not dependent on C/O but on the stability of feldspar which cannot coexist with nepheline. The relative stability of olivine versus orthopyroxene is primarily governed by the system's Mg/Si ratio. In systems exhibiting elevated Mg/Si ratios (e.g., HD24633, Mg/Si = 1.59), olivine predominates; conversely, both olivine and orthopyroxene are stable in systems with moderate Mg/Si ratios, such as HD94151 (Mg/Si = 1.11) and Lodders03 (Mg/Si = 1.02). 
Here, we also report the Mg\# ($\ce{MgO}/(\ce{MgO}+\ce{FeO})$) at 500~K across all three systems. In Lodders03, the Mg\# at 500~K is 0.69, while in HD94151 and HD24633 it reaches 0.77 and 0.96, respectively. These values reflect both the fraction of oxidised Fe available for incorporation into the olivine solid solution and the Mg/Fe ratio of the host star.

In both transitional and reduced sequences, the main change is that silicate condensation occurs at lower temperatures, since at higher temperatures oxides and silicates are replaced by metals, carbides, nitrides and silicides [see below].

\subsubsection{Metals and silicides}\label{sec:metal}

\begin{figure*}
    \centering
    \includegraphics[width=1\linewidth]{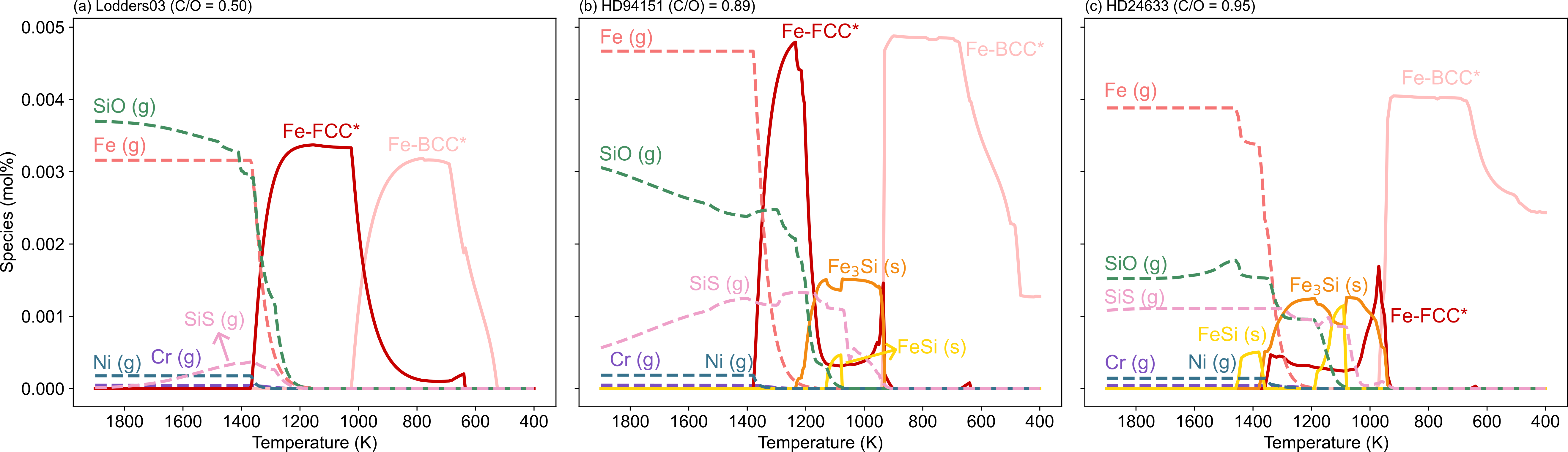}
    \caption{Same as \autoref{fig:refractory_comp_linear}, but for condensation of major metal and silicide phases.}
    \label{fig:iron_comp_linear}
\end{figure*}

\autoref{fig:iron_comp_linear}(a) shows that in \textbf{solar-like sequences}, iron primarily condenses as a metallic alloy, $\ce{Fe
_{(s)}}$ with a face-centred cubic (\textit{fcc}) structure (\ce{Fe} $f_c$ peak: 1350~K) via the reaction:
\begin{equation}
\ce{Fe_{(g)}} = \ce{Fe{^{fcc}_{(s)}}}
\label{eq:Fe_condensation}
\end{equation}
At temperatures below 1130~K, it transforms via a solid-state reaction into $\ce{Fe_{(s)}}$ with a body-centred cubic (\textit{bcc}) structure.

$\ce{Ni_{(s)}}$ and $\ce{Cr_{(s)}}$ primarily condense into solid Fe-alloy solutions from their gaseous forms, $\ce{Ni_{(g)}}$ and $\ce{Cr_{(g)}}$. Their $f_c$ peaks occur at $1360~$K for Ni and $1325~$ K for Cr, respectively. In \autoref{fig:condensation_solidsoln} we highlight the variation in $T_c^{50\%}$ values for different elements when condensation is modelled with and without the incorporation of solid solutions. Notably, the largest differences are observed for Ni and Cr ($\sim$100~K higher when solid solutions are included), since their condensation into the Fe \textit{fcc} alloy phase as dilute components is more energetically favourable than condensing pure metallic Cr$_{(s)}$ or Ni$_{(s)}$.

\autoref{fig:iron_comp_linear}(b) shows that, in \textbf{transitional sequences}, iron initially condenses as $\ce{Fe_(s)}$ (\ce{Fe} $f_c$ peak: 1365~K) via \cref{eq:Fe_condensation}. Elements \ce{Cr} and \ce{Ni} are incorporated into the iron alloy phase ($\ce{Fe_{(s)}}$), similar to solar-like sequences. Cr has $f_c$ peak at 1340~K and Ni at 1370~K. 
Eventually, $\ce{Fe_{(s)}}$ reacts with gaseous $\ce{SiO_{(g)}}$ to form $\ce{Fe3Si_{(s)}}$ (\ce{Si} $f_c$ peak: 1230~K) through the net reaction:
\begin{equation}
\ce{2SiO_{(g)} + 6Fe_{(s)} -> 2Fe3Si_{(s)} + O2_{(g)}}
\label{eq:Fe3Si_reaction}
\end{equation}
At cooler temperatures, $\ce{FeSi_{(s)}}$ consumes further \ce{SiO_{(g)}} and $\ce{Fe3Si_{(s)}}$ at $\sim$1150~K by:
\begin{equation}
\ce{Fe3Si_{(s)} + 2SiO(g) -> 3FeSi_{(s)} + O_2(g)}
\label{eq:FeSi_stabilization}
\end{equation}

Iron silicides remain stable only down to $\sim930$~K. Below this temperature, Fe is stable as Fe-$\textit{bcc}$ alloy, whereas the Si is incorporated into feldspar solid solution, pyroxene and olivine (\autoref{fig:silicate_comp_linear}(b)).

In \textbf{reduced sequences}, iron condenses as $\ce{FeSi_{(s)}}$ (\ce{Fe} and \ce{Si} $f_c$ peak: 1445~K) through the  reaction of $\ce{Fe_{(g)}}$ and $\ce{SiO_{(g)}}$ with silicon carbide:
\begin{equation}
\ce{2Fe_{(g)} + SiO_{(g)} + \ce{SiC_{(s)}} -> 2FeSi_{(s)} + CO_{(g)}}
\label{eq:FeSi_direct_condensation}
\end{equation}

The presence of $\ce{SiC_{(s)}}$ in the refractory region of the disk ($> 1400~$K), which is involved in \autoref{eq:FeSi_direct_condensation} is discussed in section \ref{sec:carbon}.

As the sequence progresses, additional reactions with $\ce{FeSi_{(s)}}$, $\ce{SiO_{(g)}}$ and $\ce{Fe_{(g)}}$ lead to the formation of $\ce{Fe3Si_{(s)}}$ (\ce{Fe} $f_c$ peak: 1370~K), and the condensation of additional $\ce{FeSi_{(s)}}$ via reaction \ref{eq:FeSi_stabilization}, and finally the breakdown of $\ce{Fe3Si_{(s)}}$ into $\ce{Mg2SiO4_{(s)}}$ and $\ce{Fe_{(s)}}$ in \textit{fcc} structure (\ce{Fe} $f_c$ peak: 1335~K) coincident with the condensation of $\ce{Mg_{(g)}}$. Cr and Ni condense into the iron alloy in accordance with the solar-like and transitional sequences.

Hence, iron silicides ($\ce{Fe_{x}Si_{(s)}}$), which are absent in solar-like systems, act as the dominant carriers of Si and Fe in systems with elevated C/O ratios at intermediate temperatures ($\sim 1400$--$1000$~K). The formation of these phases limits the amount of $\ce{Si}$ available to form $\ce{SiC_{(s)}}$ in high C/O sequences. This effect is shown in \autoref{fig:FexSi_comparision}, which compares Gibbs energy minimisation with and without iron silicides. Thus, computations that ignore $\ce{Fe_{x}Si_{(s)}}$ as a stable phase grossly overpredict the amount of C that can condense in reducing, high C/O sequences.

\subsubsection{Carbides and nitrides}\label{sec:carbon}

\begin{figure*}
    \centering
    \includegraphics[width=1\linewidth]{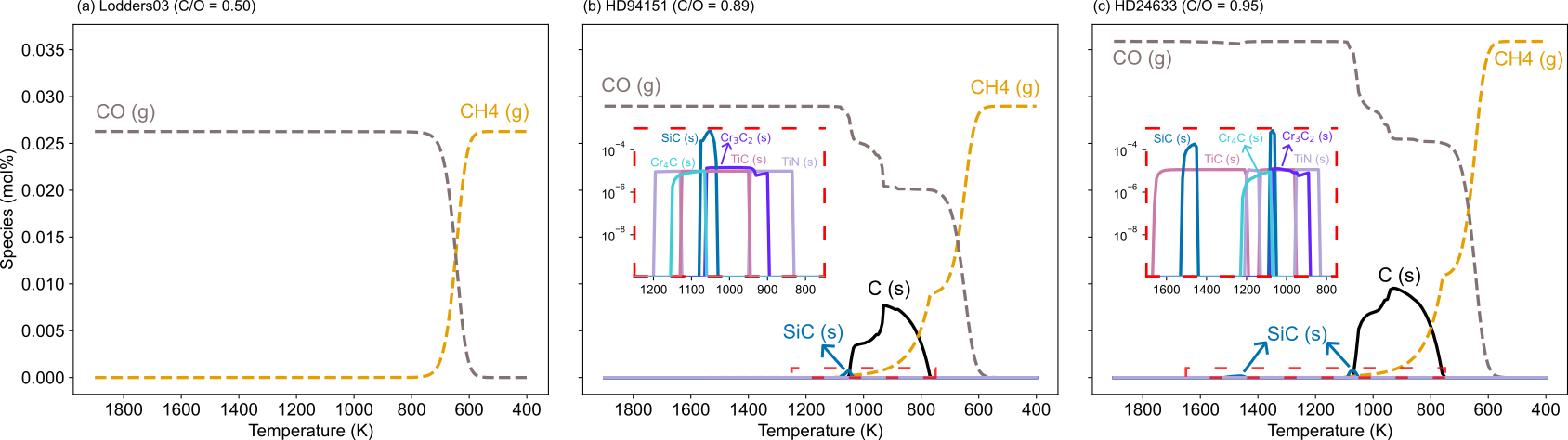}
    \caption{Same as \autoref{fig:refractory_comp_linear}, but for condensation of carbide and nitride phases. Inset presents the minor C-bearing phases in log scale.}
    \label{fig:carbon_comp_linear}
\end{figure*}

As seen in \autoref{fig:carbon_comp_linear}(a), in solar-like sequences, carbon remains in the gas form at all temperatures $> 400~K$. At $\sim$650~K and 10$^{-4}$ bar, the mole fractions of $\ce{CO_{(g)}}$ and $\ce{CH4_{(g)}}$ become equal. Below this temperature, $\ce{CH4_{(g)}}$ becomes the predominant carbon-bearing gas, facilitated by the reaction:
\begin{equation}
\ce{2CO_{(g)} + 4H2_{(g)} -> 2CH4_{(g)} + O2_{(g)}}
\label{eq:CH4}
\end{equation}

We note, however, that the kinetics of this reaction are likely too slow for equilibrium to have been reached during typical nebular lifetimes \citep{Fegley2000KineticsNebula}. 

\autoref{fig:carbon_comp_linear}(b) demonstrates that transitional sequences result in the condensation of  $\ce{SiC_{(s)}}$ followed by $\ce{C_{(s)}}$, both of which are absent in solar-like sequences. The former condenses at $\sim$1100~K via heterogeneous reactions in which some Si from previously formed $\ce{FeSi_{(s)}}$ is consumed: 
\begin{equation}
    \ce{FeSi_{(s)}} + \ce{CO_{(g)}} = \ce{Fe{_{(s)}^{fcc}}} + \ce{SiC_{(s)}} + \frac{1}{2}\ce{O2_{(g)}}
\label{eq:SiC_solidstate_cond}
\end{equation}
\ce{SiC_{(s)}} abundances are vanishingly low, of the order of $10^{-4}$ mol\%, due to iron silicide condensation, as described previously and illustrated in \autoref{fig:FexSi_comparision}. 

Graphite is the most abundant C-bearing phase, and condenses from $\sim$1050~K (\ce{C} $f_c$ peak: 930~K) via the net reaction:
\begin{equation}
\ce{2CO_{(g)} -> 2C_{(s)} + O2_{(g)}}
\label{eq:C_condensation}
\end{equation}

A distinctive feature is the emergence of $\ce{TiN_{(s)}}$ (\ce{N} $f_c$ peak: 1195~K), which condenses via the reaction of $\ce{N2_{(g)}}$ with perovskite. $\ce{TiN_{(s)}}$ transitions to $\ce{TiC_{(s)}}$ through net reaction:
\begin{equation}
\ce{2 TiN_{(s)} + 2 CO_{(g)} -> 2 TiC_{(s)} + N2_{(g)} + O2_{(g)}}
\label{eq:TiC_transitional}
\end{equation}. 
The conversion of $\ce{TiN_{(s)}}$ to $\ce{TiC_{(s)}}$, and evaporation of nitrogen, followed by subsequent recondensation of $\ce{TiN_{(s)}}$ is shown in \autoref{fig:nitrogen_condensation_comparision}.

Minor carbide phases such as $\ce{Cr4C_{(s)}}$ (\ce{Cr} $f_c$ peak: 1220~K) and $\ce{Cr3C2_{(s)}}$ occur as $\ce{Cr_{(s)}}$ dissolved in iron alloy reacts with $\ce{CO_{(g)}}$ via reactions: 

\begin{equation} 
\ce{8Cr_{(s)} + 2CO_{(g)} -> 2Cr4C_{(s)} + O2_{(g)}} \label{eq:Cr4C_formation} 
\end{equation} 

\begin{equation} 
\ce{3Cr_{(s)} + 2CO_{(g)} -> Cr3C2_{(s)} + O2_{(g)}} \label{eq:Cr3C2_formation} 
\end{equation}

\autoref{fig:carbon_comp_linear}(c) illustrates that reduced sequences display condensed phases similar to transitional sequences, but differ in two particular respects:

\begin{enumerate}
    \item $\ce{TiC_{(s)}}$ (\ce{Ti} $f_c$ peak: 1655~K) condenses directly from gaseous $\ce{Ti_{(g)}}$ via \autoref{eq:TiC_reduced}, replacing corundum as the most refractory phase. 

    \begin{equation}
    \ce{2Ti + 2CO_{(g)} -> 2TiC_{(s)} + O2_{(g)}}
    \label{eq:TiC_reduced}
    \end{equation}
    
    $\ce{TiN_{(s)}}$ here forms via solid state reaction sequentially from $\ce{TiC_{(s)}}$. This is followed by the $\ce{TiN_{(s)}}$ - $\ce{TiC_{(s)}}$ conversion dynamic as was observed in transitional sequences (\autoref{fig:nitrogen_condensation_comparision}).
    
    \item Direct condensation of $\ce{SiC_{(s)}}$ at $\sim$1510~K (refractory region) from the gas phase through the reaction:
    \begin{equation}
    \ce{Si_{(g)} + CO_{(g)} -> SiC_{(s)} + SiO_{(g)}}
    \label{eq:SiC}
    \end{equation}
    \ce{SiC} subsequently disintegrates at $\sim$1470~K such that the Si enters \ce{FeSi_(s)} and C forms \ce{CO_{(g)}} via \autoref{eq:FeSi_direct_condensation}.
    SiC condenses again at $\sim$1085~K from \autoref{eq:SiC_solidstate_cond}, similar to the condensation we observe in transitional sequences. Graphite condenses via \autoref{eq:C_condensation} (\ce{C} $f_c$ peak: 1060), $\ce{Cr4C_{(s)}}$ forms via \autoref{eq:Cr4C_formation}, and $\ce{Cr3C2_{(s)}}$ forms via \autoref{eq:Cr3C2_formation}.
\end{enumerate}

Furthermore, although not present in HD94151 $([Fe/H] = 0.04)$ and HD24633 $([Fe/H] = -0.04)$, in both transitional and reduced sequences, at elevated $[Fe/H]$, silicon oxynitride (SiNO*) phases  - mainly sinoite - condense around $\sim 1000~\mathrm{K}$ but do so over a very narrow temperature range between 900 to 800~K at vanishingly low abundances (order of $10^{-4}$ mol\%).

\subsubsection{Sulfide phases}\label{sec:sulfide}
\begin{figure*}
    \centering
    \includegraphics[width=1\linewidth]{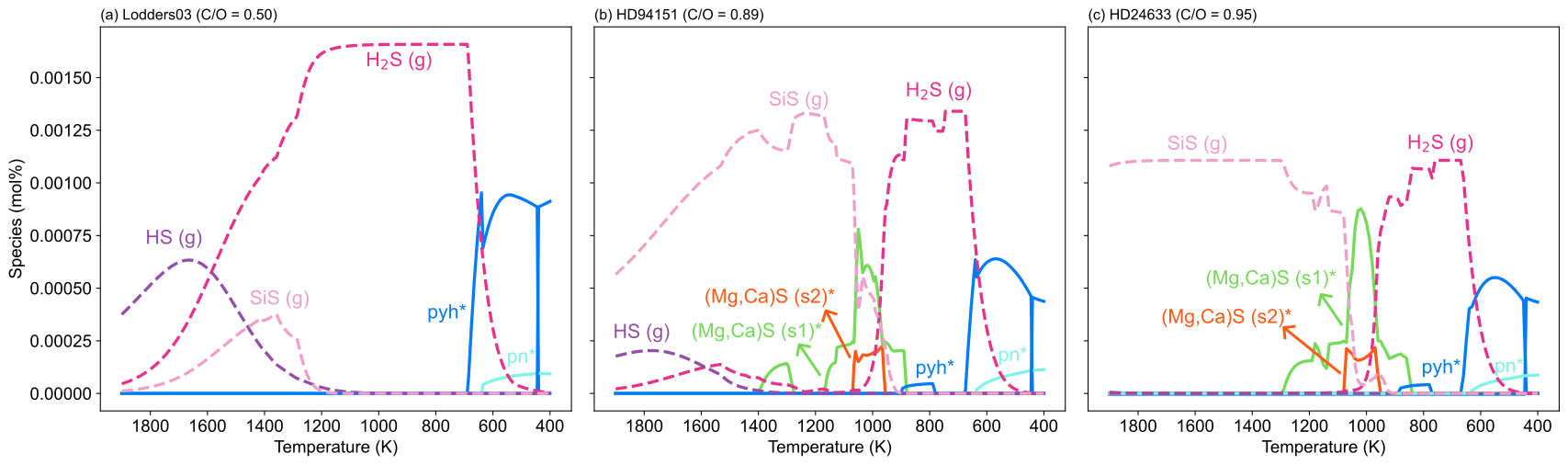}
    \caption{Same as \autoref{fig:refractory_comp_linear}, but for condensation of sulfide phases.}
    \label{fig:sulfide_comp_linear}
\end{figure*}

As observed in \autoref{fig:sulfide_comp_linear}, in solar-like sequences, sulfur is predominantly present in the gas phase, with $\ce{H2S_{(g)}}$ being the major sulfur-bearing species, followed by $\ce{HS_{(g)}}$ and $\ce{SiS_{(g)}}$. At temperatures above 700~K, sulfur remains in the gas phase before condensing directly as FeS - pyrrhotite (pyh*) with \ce{S} $f_c$ peak: 680~K, which also incorporates Ni (\autoref{fig:sulfur_solidsoln_comp_linear} via the reaction:
\begin{equation}
\ce{(Fe,Ni)_{(s)} + H2S_{(g)} -> (Fe,Ni)S_{(s)} + H2_{(g)}}
\label{eq:FeS_reaction}
\end{equation}

At lower temperatures, smaller amounts of $(\ce{(Fe,Ni)_{9}S_8})$ - pentlandite (pn*)stabilises. 

In \textbf{transitional sequences}, the lower oxygen fugacity leads to a significant fraction of silicon being sequestered in the gas phase as $\ce{SiS_{(g)}}$ as opposed to $\ce{SiO_{(g)}}$. The higher partial pressure of $\ce{SiS_{(g)}}$ compared to the solar-like sequences results in the condensation of $\ce{CaS_{(s)}}$ (\ce{Ca} $f_c$ peak: 1395~K) before it re-evaporates at $\sim$1280~K and $\ce{MgS_{(s)}}$ (\ce{S} $f_c$ peak: 1125~K) between 1400~K and 800~K. As illustrated in \autoref{fig:sulfur_solidsoln_comp_linear}, $\ce{CaS_{(s)}}$ and $\ce{MgS_{(s)}}$ form through two distinct reactions: condensation directly from the gas phase (s1) and via a heterogeneous gas-solid reaction (s2), where s1 and s2 are the compositions of two immiscible CaS-MgS solid solution phases that coexist within the miscibility gap below ~1100°C.

At lower temperatures, the partial pressure of $\ce{H2S_{(g)}}$ increases, facilitating the formation of pyrrhotite (\ce{S} $f_c$ peak: 670~K) via  \autoref{eq:FeS_reaction}, pentlandite and ($\ce{(Fe,Ni)9S8_{(s)}}$). These phases dominate the sulfur chemistry below 680~K, as per solar-like sequences.

Sulfide phases in \textbf{reduced sequences} are the same as transitional sequences: namely, condensing in sequence $\ce{CaS_{(s)}}$ (\ce{Ca} $f_c$ peak: 1290~K), $\ce{MgS_{(s)}}$ (\ce{S} $f_c$ peak: 1180~K), $\ce{FeS_{(s)}}$ (\ce{S} $f_c$ peak: 655~K), ($\ce{(Fe,Ni)9S8_{(s)}}$).

\subsection{Condensate mineralogies with varying pressure} \label{sec:disk_chem_pressure}
As the total pressure at which the computations were performed increases, the partial pressures of stable gas species tend to increase following Dalton's Law. This leads to a systematic increase in the $T_c^{50\%}$ temperatures of most elements as a function of $P$, notably those with condensation reactions depending on $p$M (where M is a metal), such as \ref{eq:Fe_condensation}. ~\autoref{tab:50-cond-temp} lists the $T_c^{50\%}$ values for elements in three representative systems - Lodders03 (solar-like), HD94151 (transitional), and HD24633 (reduced) - evaluated at total pressures of $10^{-2}$, $10^{-4}$, and $10^{-6}$ bar, also illustrated in \autoref{fig:tc50_vs_pressure}.

\begin{table*}[!htbp]
\centering
\caption{Condensation Temperatures, $T_c^{50\%}$ (K) for elements in solar-like (Lodders03; C/O$= 0.50$), transitional (HD94151; C/O$= 0.89$), and reduced (HD24633; C/O$= 0.95$) systems at three different disk pressures.}
\label{tab:50-cond-temp}
\begin{tabular}{lccccccccccc}
\hline
Star & Pressure & Al & Ca & Cr & Fe & Mg & Na & Ni & S & Si & Ti \\
\hline
Lodders03 (C/O = 0.50) & $10^{-2}$ bar & 1795 & 1645 & 1500 & 1530 & 1485 & 1180 & 1555 & 645 & 1455 & 1765 \\
 & $10^{-4}$ bar & 1655 & 1475 & 1310 & 1330 & 1335 & 1045 & 1355 & 645 & 1320 & 1595 \\
 & $10^{-6}$ bar & 1530 & 1345 & 1160 & 1180 & 1220 & 940 & 1200 & 645 & 1205 & 1460 \\
 HD94151 (C/O = 0.89) & $10^{-2}$ bar & 1620 & 1525 & 1515 & 1545 & 1240 & 1090 & 1570 & 635 & 1315 & 1660 \\
 & $10^{-4}$ bar & 1505 & 1390 & 1320 & 1345 & 1060 & 935 & 1365 & 1055 & 1170 & 1505 \\
 & $10^{-6}$ bar & 1390 & 1270 & 1170 & 1190 & 945 & 830 & 1210 & 940 & 1055 & 1375 \\
HD24633 (C/O = 0.95) & $10^{-2}$ bar & 1490 & 1455 & 1435 & 1540 & 1220 & 1040 & 1555 & 1205 & 1315 & 1850 \\
 & $10^{-4}$ bar & 1325 & 1285 & 1245 & 1340 & 1060 & 935 & 1345 & 1065 & 1155 & 1645 \\
 & $10^{-6}$ bar & 1190 & 1145 & 1105 & 1195 & 940 & 830 & 1180 & 945 & 1040 & 1470 \\
\hline
\end{tabular}
\end{table*}

\begin{figure*}[!htbp]
    \centering
    \includegraphics[width=1\linewidth]{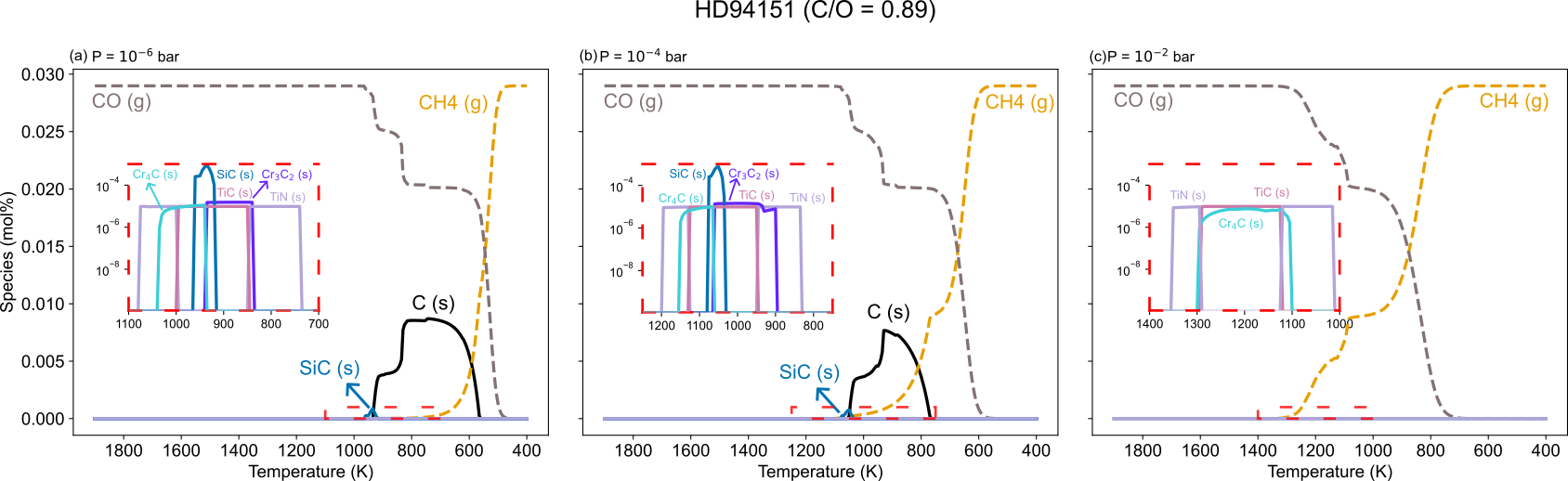}
    \caption{Stability of carbon-bearing phases in system HD94151 at pressures of(a) $10^{-6}$bar, (b)$10^{-4}$bar and (c)$10^{-2}$bar}
    \label{fig:HD94151_pres_carbon}
\end{figure*}

Exceptions to this trend arise, including, importantly, the substantial decrease in the amount of graphite and $\ce{SiC_{(s)}}$ with pressure from $10^{-6}$ to $10^{-2}$ bar (\autoref{fig:HD94151_pres_carbon} \autoref{fig:HD24633_pres_carbon}) resulting in decrease in amount of condensed C. At $10^{-2}$ bar, these species no longer condense in our simulations. This behaviour can be attributed to the pressure-dependent reaction:

\begin{equation}
\ce{CO(g) + 3H2(g) -> CH4(g) + H2O(g)}.
\label{eq:C_pressure_reaction}
\end{equation}

As pressure increases, the equilibrium shifts toward the product side due to the reduction in the number of gas molecules. Consequently, the temperature at which $x$$\ce{CO(g)}$/$x$$\ce{CH4(g)}$ = 1 shifts to higher values with increasing pressure (Fig. \ref{fig:HD94151_pres_carbon}). As $\ce{CO(g)}$ is consumed in the formation of methane, the partial pressure of $\ce{CO(g)}$ decreases below the saturation threshold required for the condensation of $\ce{C_{(s)}}$ and $\ce{SiC_{(s)}}$.

Another important observation pertains to sulfide condensates. For the solar-like sequence (e.g., the Lodders03 system), \autoref{tab:50-cond-temp} indicates that sulfur $T_c^{50\%}$ remains constant at 645~K. The condensation of $\ce{FeS_{(s)}}$ occurs via \autoref{eq:FeS_reaction}, but this reaction is pressure-independent as the number of gas molecules remain equal between the reactants and products \citep{Lewis1972LowNebula}. Since, in the solar-like sequence, $\ce{FeS_{(s)}}$ is the only sulfur-bearing condensed phase, the condensation temperature of S is independent of pressure. In contrast, for transitional and reduced sequences (e.g., HD94151 and HD24633), \autoref{tab:50-cond-temp} shows that sulfur $T_c^{50\%}$ changes with pressure. The condensation of $\ce{(Ca,Mg)S_{(s)}}$ at these temperatures is governed by a pressure-dependent reaction.

\begin{figure*}[!htbp]
    \centering
    \includegraphics[width=1\linewidth]{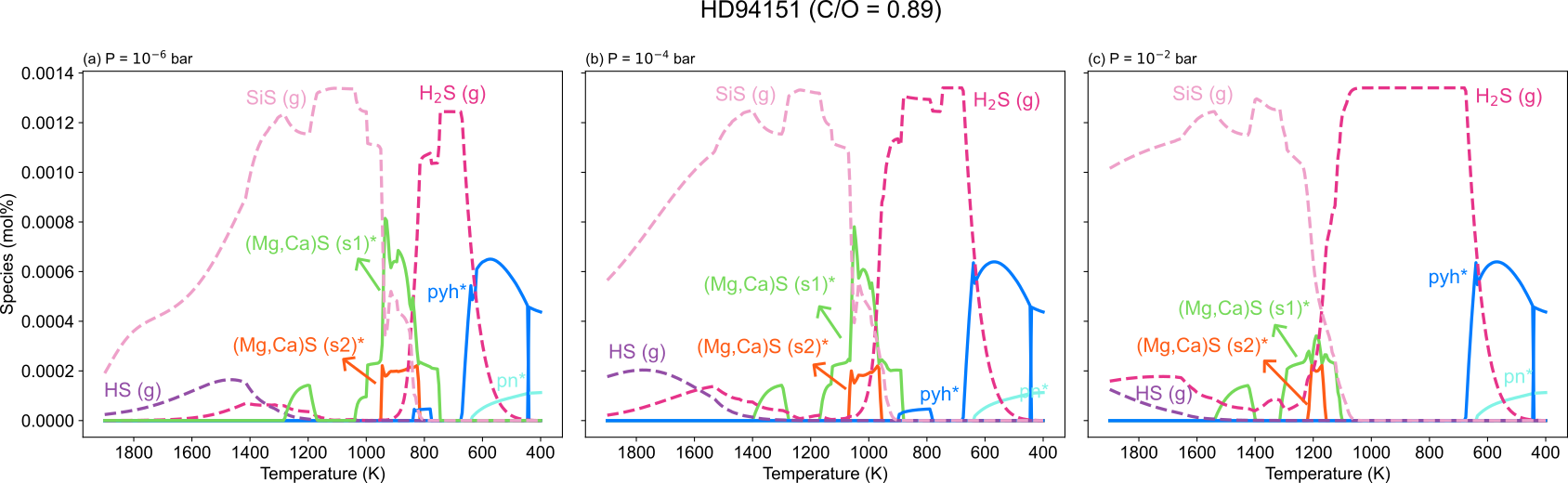}
    \caption{Same as \autoref{fig:HD94151_pres_carbon}, but for condensation of sulfide phases.}
    \label{fig:HD94151_pres_sulfur}
\end{figure*}

Furthermore, although the condensation of $\ce{(Ca,Mg)S_{(s)}}$ in both transitional and reduced sequences shifts towards higher temperatures with increasing pressure, the total amount of condensed $\ce{(Ca,Mg)S_{(s)}}$ decreases. This occurs because condensation is controlled by the partial pressure of $\ce{SiS_{(g)}}$, which is progressively converted to $\ce{H2S_{(g)}}$ at higher temperatures under elevated pressures.

\subsection{Bulk compositions of hypothetical planet(esimal)s} \label{sec:planet_composition}

As described in section \ref{sec:accretion_method}, we employ the stochastic accretion model of \cite{Sossi2022StochasticEarth} to accrete condensates within a given feeding zone defined by the inflection point $T_0$ and steepness $\sigma$ of a logistic curve. Higher $T_0$ values favour the incorporation of refractory elements that condense at high temperatures, whereas lower $T_0$ values allow greater proportions of volatile elements to accrete. It follows that larger $\sigma$ values allows integration of material across broader temperature and pressure gradients, while smaller $\sigma$ values restrict accretion of solids that condensed over a narrower temperature interval, thereby preserving localised chemical signatures tied closely to the specific $T_0$ \autoref{fig:sigma_comparision}. 

To assess the compositional changes in planet(esimals) formed through the accretion of condensates, we focus the results discussed here on the $T_0$-dependent variations in their bulk composition across three systems (Lodders03, HD94151, and HD24633) within an Earth-like feeding zone ($\sigma = 225$) at a disk pressure of $10^{-4}$~bar. We also discuss the influence of disk pressure on planetesimal composition by comparing accretion of solids in sequences at $10^{-6}$, $10^{-4}$, and $10^{-2}$~bar.

\begin{figure*}
    \centering
    \includegraphics[width=0.8\linewidth]{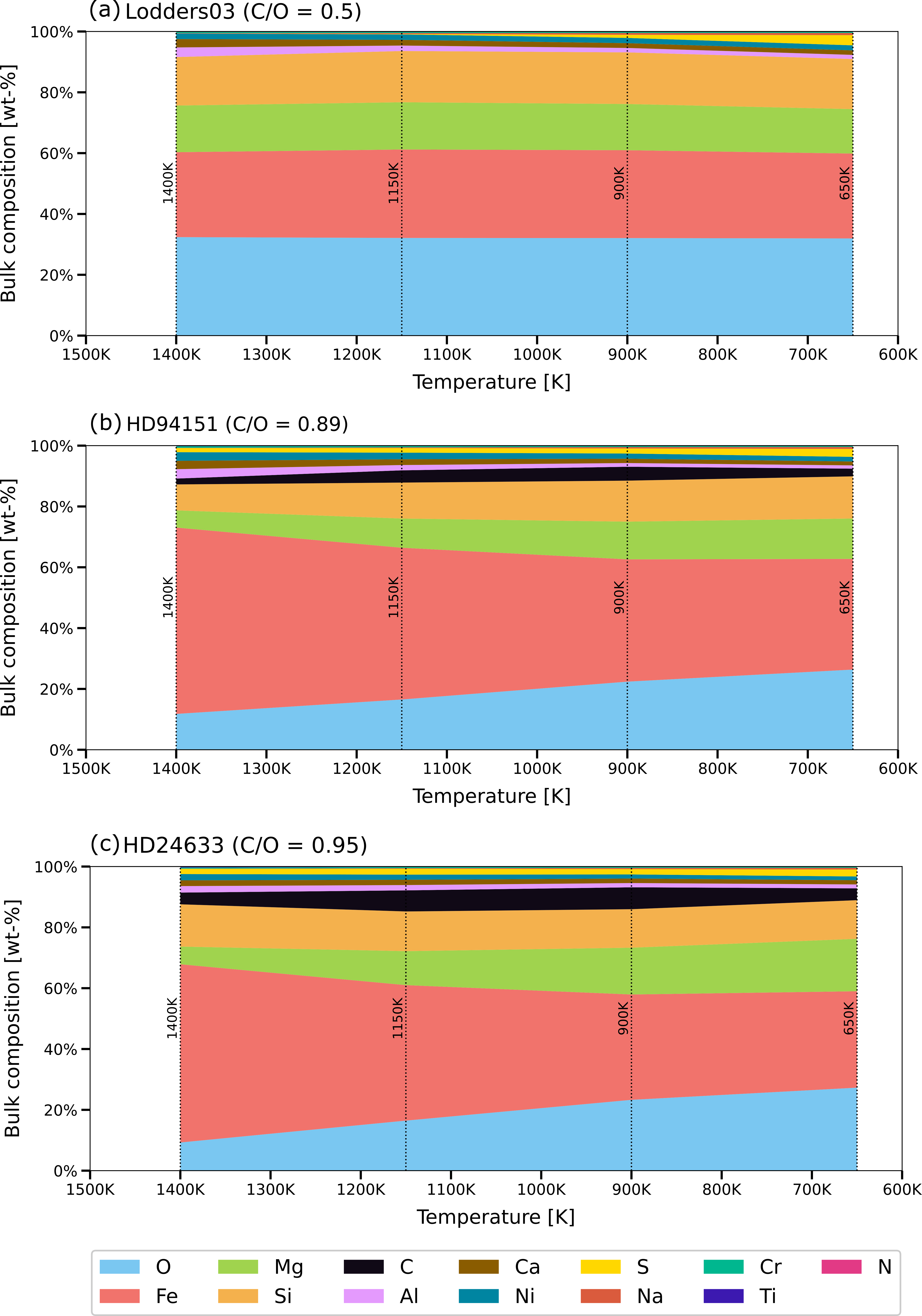}
    \caption{Bulk composition of accreted planet(esimal)s at various $T_0$ marked in dotted lines, with a feeding zone of 225~K, a) Around Lodders03; b) Around HD94151, c) Around HD24633.}
    \label{fig:t0_planet}
\end{figure*}

\begin{figure*}
    \centering
    \includegraphics[width=1\linewidth]{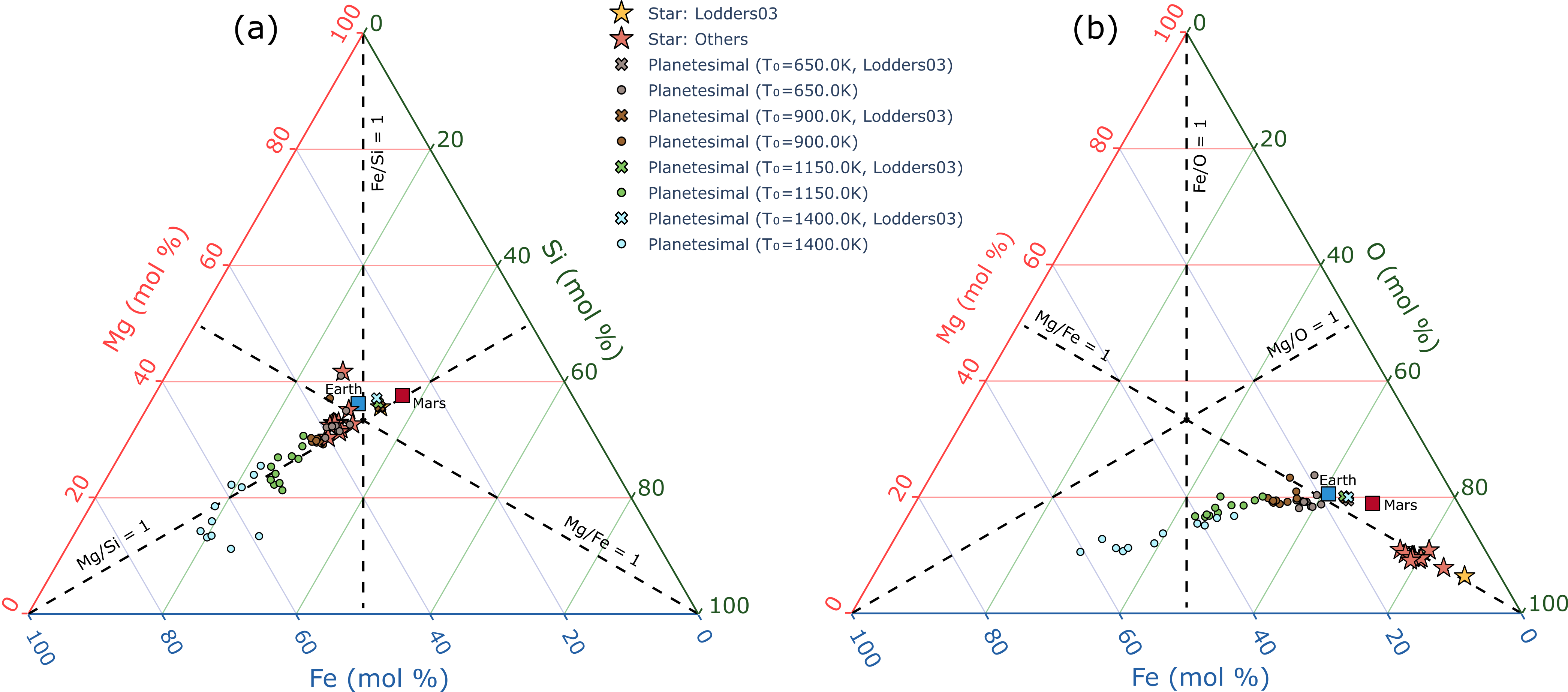}
    \caption{Ternary diagram (a) Mg-Fe-Si (mol \%) and (b) Mg-Fe-O displaying high C/O stars and modeled planet(esimals). Sun \citep{Lodders2003SolarElements}, Earth \citep{Fischer2020TheCore} and Mars \citep{Yoshizaki2020TheMars} included for reference.}
    \label{fig:Mg-Fe-Si_ternary}
\end{figure*}

Shown in \autoref{fig:t0_planet} and \autoref{fig:Mg-Fe-Si_ternary}(a), planet(esimal)s modelled based on the solar-like sequence (here Lodders03; C/O = 0.50) exhibit a narrow compositional range in molar Fe, Mg and Si abundances independent of the $T_0$ of the planet(esimal)s. 
Furthermore, their Fe/Mg, Mg/Si and Fe/O ratio are nearly indistinguishable across the temperature regimes from which they accreted. 

In contrast, for transitional and reduced systems (here HD94151 and HD24633), the highest molar Fe/Mg ratio ($\sim4-5$) and Fe/O ratio ($\sim1.4-2$) occur at $T_0$ = 1400~K, where only small fractions of the total Mg and Si budget have accreted. The Fe/Mg ratio decreases with decreasing $T_0$, approaching their host star's ratio only at $T_0 \sim 650$K (\autoref{fig:Mg-Fe-Si_ternary}(a)). The Fe/O ratio also decreases with $T_0$, reflecting progressive condensation of oxygen-bearing phases as the system cools (\autoref{fig:Mg-Fe-Si_ternary}(b)). Even planet(esimal)s formed with $T_0$ of 650~K do not converge to the oxygen content of their host star, because a large fraction of the oxygen budget remains gaseous above 400~K. This is also evident for the solar system planets shown (Earth and Mars), of which the O budget is clearly lower than that of the Sun. The Mg/Si ratio, on the other hand, increases with decreasing $T_0$, which reflects the relatively refractory nature of Si over Mg in these C/O rich disks, mainly due to its sequestration into $\ce{Fe_{x}Si_{(s)}}$ as shown in section \ref{sec:metal}. 

\begin{figure*}[!ht]
    \centering
    \includegraphics[width=0.9\linewidth]{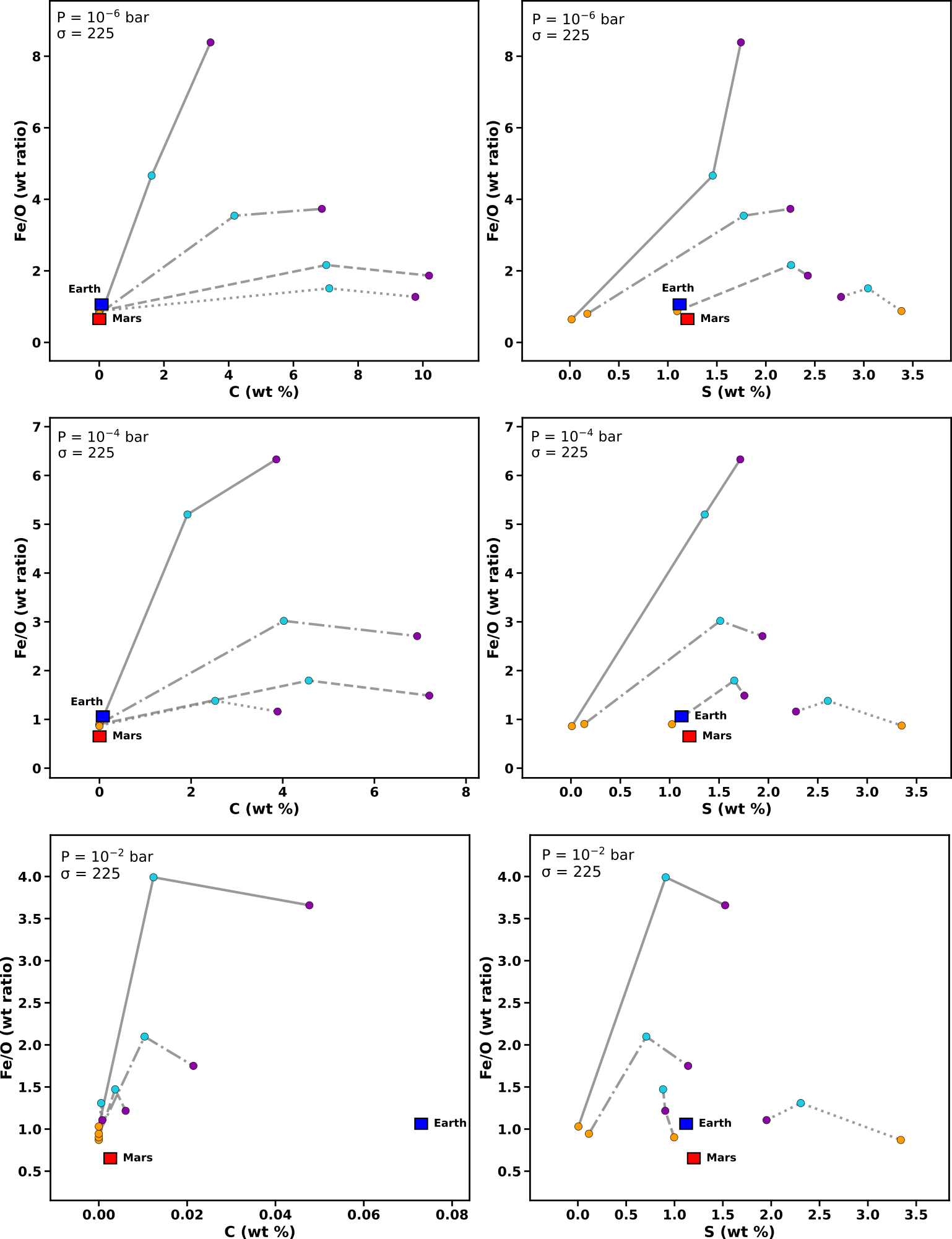}
    \caption{C (wt$\%$) and S (wt $\%$) of planet against Fe/O wt $\%$ ratio of star for disk pressures of $10^{-6}$ bar (top panel); $10^{-4}$ bar (middle panel); and $10^{-2}$ bar (bottom panel). Sun \citep{Lodders2003SolarElements}, Earth \citep{Fischer2020TheCore} and Mars \citep{Yoshizaki2020TheMars} included for reference.}
    \label{fig:S-C-wt}
\end{figure*}

As shown in \autoref{fig:S-C-wt}, the behaviour of volatile elements such as sulfur (S) and carbon (C) also varies significantly. In the solar-like system (here Lodders03), S is incorporated into planet(esimals) only at much lower temperatures ($< 700~\mathrm{K}$), with C essentially remaining volatile throughout. In contrast, in high C/O systems, S and C-bearing condensates occur at higher temperatures, leading to higher fractions of them compared to solar system planets, even at $T_0 \geq 1400~\mathrm{K}$. 

We note that while we adopt $\sigma = 225$ as an agnostic feeding zone width informed by an Earth accretion scenario, the effective feeding zone of exoplanets may vary depending on dynamical history, migration, or final planetary mass. To evaluate the sensitivity of our results to this assumption, we repeat the calculations for both narrower ($\sigma = 50$) and broader ($\sigma = 300$) feeding zones, as shown in \autoref{fig:sigma_comparision}. The systematically more marked variations in Fe/O, Fe/Mg, and Mg/Si ratios, together with the enhanced S and C abundances highlighted above, persist across these cases. Our conclusions are therefore not contingent on the specific choice of feeding zone width.

\begin{table*}[htbp]
\centering
\caption{Bulk composition (wt\%) of planet(esimal)s forming at $T_0 = 1150$~K from an Earth-sized feeding zone ($\sigma = 225$ K) for systems Lodders03 (C/O = 0.50), HD94151 (C/O = 0.89), and HD24633 (C/O = 0.95).}
\label{tab:bulk_comp_wt}
\begin{tabular}{lccc}
\hline
Element & Lodders03 (C/O = 0.50) & HD94151 (C/O = 0.89) & HD24633 (C/O = 0.95) \\
\hline
Ni  & 1.8 & 2.18 & 1.68 \\
Fe  & 29.09 & 49.899 & 44.559 \\
Cr  & 0.395 & 0.46 & 0.383 \\
Ti  & 0.09 & 0.106 & 0.138 \\
Ca  & 1.87 & 1.89 & 1.75 \\
S   & 0.135 & 1.51 & 1.94 \\
Si  & 16.85 & 11.83 & 13.04 \\
Al  & 1.78 & 1.74 & 1.75 \\
Mg  & 15.54 & 9.61 & 11.18 \\
Na  & 0.338 & 0.22 & 0.187 \\
O   & 32.11 & 16.53 & 16.46 \\
C   & 0 & 4.02 & 6.93 \\
He  & 0 & 0 & 0 \\
H   & 0 & 0 & 0 \\
N   & 0 & 0.0074 & 0.0010 \\
\hline
\end{tabular}
\end{table*}

\autoref{tab:bulk_comp_wt} shows the bulk compositions of planet(esimal)s forming at $T_0 = 1150$~K ($T_0$ assumed for Earth; \citet{Sossi2022StochasticEarth}). Relative to the solar-like sequence, bodies forming in C/O-enriched systems are markedly enriched in Fe and Ti, while depleted in Mg, O, Si, and Na. Refractory elements such as Ca and Al accrete in comparable amounts across all systems, whereas volatile elements, including S, C, and N - which condense only minimally (or not at all) above 400~K in solar-like sequences-contribute up to a few wt\% in high C/O systems.

The variability in these trends is further influenced by total pressure. \autoref{fig:S-C-wt} shows that for a given system, planet(esimal)s formed at $10^{-2}$ bar contain up to $\sim100\times$ lower carbon mass fractions relative to those formed at $10^{-6}$ bar. In contrast, an increase in pressure leads to condensation of sulfur-bearing $\ce{CaS_{(s)}}$ and $\ce{MgS_{(s)}}$ at increasingly higher temperatures in transitional and reduced systems. However, because these phases condense and re-evaporate over a relatively narrow temperature interval, their influence on the corresponding S budget of a hypothetical planet(-esimal)s is limited. In the solar-like system, sulfur condenses only via the pressure-independent reaction, resulting in the formation of troilite at $\sim$680~K, and hence, its incorporation into planet(esimals) is roughly independent of pressure. 
These calculations highlight the contrasting effects of pressure on carbon and sulfur overall: higher pressures suppress the amount of carbon-bearing condensates but have minimal impact on that of sulfur-bearing condensates, regardless of the system's C/O ratio.
 
\section{Discussion} \label{sec:discussion}
\subsection{Disk chemistry and redox conditions}

As detailed in section \ref{sec:Results}, we identify three distinct regimes of disk chemistry governed by nebular carbon-to-oxygen (C/O) ratios. For $\mathrm{C/O} \lesssim 0.7$, condensation follows a solar-like sequence dominated by silicates and oxides. In the transitional regime ($0.7 \lesssim \mathrm{C/O} \lesssim 0.91$), reduced species appear alongside these common phases. At $\mathrm{C/O} > 0.92$, reduced condensates such as \ce{TiC_{(s)}} and \ce{SiC_{(s)}} dominate the high-temperature chemistry, with a corresponding reduction in silicate and oxide stability.  

This classification aligns with the regime boundaries of \citet{Shakespeare2025TheStars}, who distinguish silicate-dominant, intermediate, and carbide-dominant systems in a dynamic disk framework. Our results demonstrate that these compositional regimes can also be reproduced with equilibrium condensation alone. Transitions between them are fundamentally controlled by the disk’s oxygen budget relative to other elements (\autoref{fig:delta_iw_plot_final}).  

\citet{Adams2025EffectsStars} investigated presolar grain formation in AGB outflows of carbon stars, focusing on how carbon chemistry governs the condensation of \ce{TiC_{(s)}}, \ce{C_{(s)}}, and \ce{SiC_{(s)}}. In their models, where $\mathrm{C/O} > 1$, graphite condenses over $\sim2200$--1100~K through pathways such as
\begin{equation}
\ce{C_{(g)}} \rightarrow \ce{C_{(s)}}
\label{R1_adams}
\end{equation}
and
\begin{equation}
\ce{C2H2_{(g)}} \rightarrow 2\ce{C_{(s)}} + \ce{H2_{(g)}}.
\end{equation}
enabling it to condense up to very high temperatures (2200~K). In contrast, in our stellar sample with $\mathrm{C/O} < 1$, graphite does not condense via these mechanisms, since nearly all carbon remains locked in \ce{CO_{(g)}} and the partial pressures of \ce{C_{(g)}} and \ce{C2H2_{(g)}} are negligible. Therefore, graphite condenses only by reaction \ref{eq:C_condensation} at modest temperatures ($\sim$1000~K at 10$^{-4}$ bar).  

Both their results and those reported here emphasise that, in addition to C/O, total pressure regulates gas chemistry. This regulation is non-linear: not all condensation temperatures ($T_c^{50\%}$) simply shift uniformly with pressure (as expected from Dalton's Law and the Clausius-Clapeyron equation) but are mediated by specific gas-phase reactions. Notably, increasing pressure stabilises $\ce{CH4_{(g)}}$ relative to $\ce{CO_{(g)}}$, suppressing the formation of $\ce{SiC_{(s)}}$ and $\ce{C_{(s)}}$ (section \ref{sec:disk_chem_pressure}). This mechanistic link explains the findings of \citet{Bond2010THESIMULATIONS}, who noted that even moderately carbon-rich systems (e.g., GI777, $\mathrm{C/O} = 0.78$) develop carbon-dominated chemistries when modelled at very low pressure ($10^{-5}$~bar).  

Together, these insights highlight that while C/O is the primary control on condensation regimes, pressure and metallicity exert important influences on the stability and abundance of condensates.
 
\subsection{Star-planet chemical connection: refractory and volatile elements}

Our results show that condensation temperatures derived for the solar nebula are not reliable proxies for element volatility in non-solar-like systems, consistent with the conclusions of \citet{Timmermann2023RevisitingCompositions} and the recent work of \citet{Spaargaren2025Proto-planetaryFormation}. As shown in \autoref{tab:fc-temperatures}, solar-like sequences yield single, well-defined $T_c^{50\%}$ values for most elements (except O, see below), whereas transitional and reduced sequences often result in condensation and re-evaporation phenomena, making the definition of a single $T_c^{50\%}$ for a given element a poor descriptor of its volatility (e.g., S, C and Mg). Moreover, while the $T_c^{50\%}$ of major rock-forming elements such as Mg, Ca, and Al decreases with increasing C/O, metallic elements such as Fe and Ni are comparatively insensitive-except where they are sequestered into silicides. Reduced phases including \ce{TiC_{(s)}} and \ce{TiN_{(s)}} render Ti, C and N more refractory.  

Another simplification arises in the assumption that refractory element ratios (Fe/Si, Mg/Si) in planets reflect those of their host stars \citep{Thiabaud2015ElementalPlanets, Jorge2022FormingComposition}. While this 1:1 correspondence holds in solar-like disks-where refractory species condense fully at high temperature \citep{Moriarty2014CHEMISTRYCOMPOSITION} - our models show significant divergence in transitional and reduced sequences. As illustrated in \autoref{fig:Mg-Fe-Si_ternary}, planetary Fe/Si and Mg/Si ratios deviate strongly from stellar values at higher $T_0$, converging only at lower $T_0$.

A similar increase in Fe/O with increasing condensation temperature is also seen in \citet{Jorge2022FormingComposition}. In their case, however, the enrichment primarily reflects the intrinsically higher refractory nature of Fe relative to O under the assumption of in situ planetesimal formation without radial mixing. If material in their disk were radially mixed over a finite feeding zone, this local Fe enrichment would be averaged out. In contrast, in our reduced and high C/O sequences, condensation of oxygen-bearing minerals is delayed to significantly lower temperatures. As a result, even with radial mixing, Fe-rich solids are incorporated prior to substantial oxide and silicate condensation, amplifying Fe/O variability with $T_0$. Because the $T_0$ of a given exoplanet cannot be predicted \textit{a priori}, this adds a degree of freedom to the inference of planetary compositions from stellar abundances alone. Moreover, there is no mandate that the C/O ratio of the region of the disk from which the planet formed reflects that of the bulk star, as evidenced by the variety in the chemistry of chondritic meteorites, and in particular the enstatite chondrites \citep{Keil1968MineralogicalChondrites, Ebel2011EquilibriumOrigins}. Consequently, using host star compositions as direct proxies for that of the planet is tenuous.

\subsection{Implications for bulk planet compositions}

Since our study does not explicitly adopt a physical disk model [see section \ref{sec:limitations} for further information], direct comparison with previous works-such as \citet{Bond2010THESIMULATIONS}, \citet{Moriarty2014CHEMISTRYCOMPOSITION}, \citet{Shakespeare2025TheStars}, and \citet{Spaargaren2025Proto-planetaryFormation} is limited. However, a common outcome across these studies is the predicted compositional gradient from silicate-dominated to carbon-enriched or carbide-dominated planet(esimals) as a function of increasing disk C/O ratio. In our dataset, the highest C/O value is associated with the star HD24633 (C/O = 0.95). At $\sigma = 225 K$, the planetesimal formed at $T_0 = \text{900}$~K in this system exhibits the highest carbon abundance, with carbon reaching nearly 10~wt\% - a value consistent with the predictions of \citet{Bond2010THESIMULATIONS}, who report carbon enrichment up to 9.8~wt\% for planet forming in disks with C/O between 0.8 and 1.

\citet{Shakespeare2025TheStars} further emphasised that planetesimal compositions in high-C/O disks are more sensitive to their orbital location than those in lower-C/O environments. This sensitivity is echoed in our results: as shown in section \ref{sec:planet_composition}, the variation in bulk composition as a function of $T_0$ is modest ($\leq$4~wt\%) in solar-like sequences. However, in transitional and reduced regimes, particularly for major rock-forming elements such as Fe, Si, Mg, and O, this variability increases significantly depending on $T_0$ \autoref{fig:t0_planet}. 

In solar-like systems, sulfur condenses below 700 K, hence its abundance in planet(esimals) is strongly dependent on $T_0$, and coincides with the temperature at which already condensed iron is oxidised from metal alloy to form troilite and fayalite, meaning that S mole fraction in the bulk condensate is anti-correlated with $T_0$ [see also \citet{Sossi2025PhysicochemicalPlanets}]. On the other hand, C and S are present at all $T_0$ in these high-C/O systems \autoref{fig:S-C-wt}, suggesting that even planet(esimals) forming closer to the host star can acquire substantial volatile inventories-a result that departs from expectations based on solar-nebula analogues.

\citet{Spaargaren2025Proto-planetaryFormation} use the devolatilisation trend defined for the Earth-Sun \citep{Wang2019TheAbundances} to deduce distinct planetary populations. Compared to their host star abundances, planets formed in disks with C/O $> 1.04$ are depleted in Mg (i.e., Mg$_{planet}$/Mg$_{star} < 1$), those between 0.84-1.04 are depleted in both Mg and Si, whereas broadly solar-like sequences with C/O $\leq$ 0.84 have planets with roughly solar (stellar) Fe, Mg and Si relative abundances. Compared to the approach of \cite{Spaargaren2025Proto-planetaryFormation} that uses $T_c^{50\%}$ and thus cannot accurately account for the abundances of O, C and S in accreting material, we modify the value of $T_0$. Consequently, we find that in reduced disks, a large range of planetary compositions can also form as a function of $T_0$ in a single system of fixed C/O. For example, \autoref{fig:t0_planet}(b, c) show Mg, Si-depleted planets at high $T_0$, contrasting with planets that have Mg, Fe, Si in the same relative abundances as those in the star at low $T_0$ of 650~K. 

Although we do not model this explicitly, it is worth noting that pressure and temperature are not static parameters and evolve both radially and temporally. Radially, both temperature and pressure decrease with distance from the central star, following approximate power laws in viscously heated regions and shallower gradients in irradiation-dominated outer regions \citep{Moriarty2014CHEMISTRYCOMPOSITION, Shakespeare2025TheStars}. Qualitatively, because $P$ and $T$ are positively correlated in viscous disks \citep{Dullemond2010TheDisks,Bitsch2015TheDiscs}, the condensation temperatures of most metals will increase relative to C (whose condensation in graphite and \ce{SiC} is not favoured at higher $P$ due to the stability of CH$_{4(g)}$ relative to CO$_{g}$). As a result, condensation at small heliocentric distances where $P$ is high may result in sulfur and metal-rich planets with only modest carbon contents, despite the high C/O ratios in exoplanetary systems whose planets orbit close to their host star, presuming they formed \textit{in-situ} (i.e., in the absence of Type I migration). 
Such chemically driven enrichment of metallic compounds provides a natural pathway toward forming planets with large core-mass fractions, characteristic of the growing population of Super-Mercuries inferred from mass-radius measurements \citep{Adibekyan2021CompositionStars}. In this context, \citet{Mah2023FormingAbundances} demonstrated that variations in stellar Mg/Si ratio can similarly promote iron-rich pebbles and the formation of Mercury-like planets. Our results extend this chemical framework by showing that super-Mercuries may also arise in high C/O environments. 

Temporally, as the disk cools and accretion slows, the $P$–$T$ conditions at any given radial location decline, shifting condensation fronts inwards and altering the stability of key condensates \citep{Bond2010THESIMULATIONS}. Under such conditions, C condensation reactions are promoted relative to those of metals at lower total pressures (ref. section \ref{sec:disk_chem_pressure}). 
The carbon enrichment in these systems arises through equilibrium condensation during nebular cooling. Our results show that bulk compositions enriched in C, S, and N are produced in otherwise oxidized planet(esimals) in transitional and reduced systems, specifically for formation temperatures below $\sim 900 $K (\autoref{fig:t0_planet}). This mechanism is fundamentally distinct from the carbon-rich 'soot-planet' scenario proposed by \citet{Li2026SootWorlds}, which has been put forward as an alternative explanation for the observed population of low-density sub-Neptunes in place of commonly assumed  H$_2$-He envelopes \citep{Owen2017ThePlanets}. Their scenario proposes carbon enrichment through inheritance and survival of interstellar organic material (CHON-rich dust) largely independent of the disk C/O ratio. 
Planet(esimals) forming at later, cooler stages of disk evolution in C/O-rich systems may therefore give rise to comparatively C- and S-rich planets relative to their solar-system counterparts. Quantifying the extent to which such compositional differences influence planetary bulk density will require dedicated mass–radius modelling tailored to these compositions.

\subsection{Planetary differentiation}
The partitioning behaviour of elements in planetary interiors is governed by a combination of the system's oxygen budget (and bulk composition more broadly), the intrinsic chemical potentials of the stable compounds of each of the elements (broadly classified as lithophile, chalcophile, or siderophile), and the thermodynamic stability of phases at the relevant $P-T$ conditions.  
As noted in \citet{Sossi2019EvaporationTheory}, the generalised order of oxygen consumption follows the sequence: Ca $>$ Al $>$ Mg $>$ Si $>$ Fe. Thermodynamic modelling by \citet{Unterborn2014TheHabitability} suggests that, across a wide range of pressures and temperatures relevant to planetary interiors, iron oxidises preferentially before carbon. However, subsequent experimental work paints a more equivocal picture, with C solubility in the metallic phase indicated to strongly \citep{Fischer2020TheCore} or modestly \citep{Blanchard2022TheSystem} decline with increasing pressure up to $\sim$70~GPa, and potentially with H content of the system \cite{Gaillard2022AEquilibration}. Consequently, the precise C budget of terrestrial planetary cores remains poorly constrained.

Planetary oxidation state can be crudely estimated by its \autoref{fig:oxygen_ratio_star_planet}, the $(\ce{O} - \ce{Mg} - 2\ce{Si})/\ce{Fe}$ ratio (\citep{Wang2022DetailedExoplanets}). For nearly all planet(esimals) examined in our study, this value is negative, indicating that even Si remains partially in its metallic form in most cases, with the extreme case of a planetesimal formed in system HD24633 at high $T_o = 1400~$K having about 96\% of Si unoxidised. In practice, this result is unlikely to hold because, for compositions in which there is sufficient O to produce SiO$_2$-rich mantles, its dissolution into metallic iron proceeds as:

\begin{equation}
    \ce{SiO2_{(l)}} + 2\ce{Fe_{(l)}} = \ce{Si_{(l)}}+ 2\ce{FeO_{(l)}} 
    \label{eq:Si_Fe_diss}
\end{equation}

leading to a self-limiting process in which the increasing FeO activity in the mantle stifles further partitioning of Si into the core, particularly at high pressure and temperature \citep{Gessmann2001SolubilityCore, Rubie2011HeterogeneousEarth, Siebert2013TerrestrialConditions}. However, in many reduced sequences, Si is already present in reduced form (as is Fe), meaning these planets cannot undergo auto-oxidation as implied by eq. \autoref{eq:Si_Fe_diss}. Moreover, the incorporation of oxygen into the cores of the terrestrial planets occurs by dissociation of FeO into Fe and O \citep{ONeill1998Oxide-metalCore}, the progress of which is again limited by the paucity of FeO in reduced bulk compositions. Consequently, cores of reduced planet(esimal)s will be massive (i.e., the core mass fraction is high) and rich in Si and poor in O. A feature common to all sequences is that only planet(esimals) formed at lower $T_0$ display more oxidised compositions similar to solar-like planet(esimals), as expected given the convergence of $f$O$_2$ (\autoref{fig:delta_iw_plot_final}). These results imply that, for the majority of planet(esimals) modelled here with $T_0 > 400$~K, both Fe and C are expected to remain in reduced form, with very low activity in the silicate portion of the planet. However, the metallic core may not be a single, homogeneous phase, because experimental studies have highlighted that light elements influence phase equilibria:

\begin{itemize}
    \item \textbf{Fe--S--Si system:} A liquid--liquid immiscibility gap between S-rich and Si-rich melts has been reported up to $\sim$12~GPa, above which it is expected to close \citep{Sanloup2004ClosurePressure, Siebert2004TheConditions, Morard2008InPressure, Tateno2018MeltingCore}.

    \item \textbf{Fe--C--S system:} Sulfide--carbide immiscibility is observed, with the gap closing near $\sim$5.5~GPa \citep{Corgne2008C-Planetesimals}. Ternary experiments up to $\sim$6~GPa further show that the solubility of C in Fe-rich melts decreases with added S, and that immiscible Fe-carbide-rich and Fe-sulfide-rich liquids can coexist. Above $\sim$5$-$6~GPa, however, C and S become mutually soluble, suppressing super-liquidus immiscibility in volatile-rich systems \citep{Sanloup2004ClosurePressure}.
\end{itemize}

It is important to note that most of these experimental constraints were obtained under conditions approximating terrestrial $f\ce{O2}$. In contrast, the highly reducing, volatile-rich environments expected in some exoplanetary systems may significantly alter light-element solubility ranges, shifting miscibility boundaries, or generating new phase regimes. Dedicated high-pressure experiments at low $f\ce{O2}$ are therefore essential to constrain stable phase assemblages, and C, S, and Si partitioning during core formation in such planetary contexts.
Beyond experimental constraints, additional modelling developments are required to translate bulk composition and core-mantle differentiation outcomes into observable planetary properties. In particular, robust equations-of-state are necessary to model planetary interiors self-consistently and to reliably predict the degree to which variation in mineralogy affects the mass-radius relationship for comparison with observed exoplanets \citep{Unterborn2016SCALINGMODELS, Baumeister2025FundamentalsExoplanets}. Finally, coupling interior and differentiation models with atmospheric evolution and degassing models will be required to predict the composition of secondary atmospheres expected from such reducing interiors, enabling direct comparison with current and future observations \citep{Bower2025DiversitySub-Neptunes}. 

\subsection{Limitations and outlook}\label{sec:limitations}
While we restrict this study to FGK dwarfs (with $T_{\mathrm{eff}} > 5000$~K), it is important to consider how such compositional diversity may extend to cooler stars, particularly M dwarfs, which are favourable targets for exoplanet detection via transit and radial-velocity techniques \citep{Nutzman2008DesignDwarfs, Trifonov2018TheSystems}. However, reliable elemental abundance determinations for M dwarfs remain challenging due to their faintness and strong molecular line blending \citep{Reiners2018TheStars}. Recent comparative analyses have nevertheless shown that abundance trends of major rock-forming elements in M dwarfs may differ systematically from those of GK dwarfs, suggesting that scaling M-dwarf abundances from GK-dwarf trends is not preferred \citep{Wang2024TowardsDwarfs}. Ongoing improvements in atomic and molecular line lists, atmospheric modelling \citep{Hejazi2024High-resolutionDwarf, Olander2025PerformanceModule}, and emerging data-driven approaches based on large spectroscopic surveys \citep{Behmard2025ASDSS-V} are expected to significantly improve abundance constraints, making it feasible to translate the implications of this work to the M-dwarf regime. 

A key assumption made in this study is that the system attains thermodynamic equilibrium at any $P-T$, thereby neglecting reaction kinetics, metastable phases, or the dynamical temporal evolution of the disk. This choice is motivated by both scope and tractability. Our framework isolates the first-order chemical controls - notably the C/O ratio, pressure, and metallicity - on condensation without combining them with dynamical transport effects. Imposing thermodynamic equilibrium also provides a direct means of comparison with the classical solar condensation sequence \citep{Grossman1972CondensationNebula, Lodders2003SolarElements}. 
More realistic models might consider the intrinsic kinetic inhibition to the progress of condensation reactions that become more marked down-temperature \citep{Fegley2000KineticsNebula} and lead to mineralogies distinct from those predicted at equilibrium \citep{Charnoz2026Non-equilibriumSolids}. The chemical reaction timescale must be shorter than nebular lifetime ($10^5 - 10^7$) years. As noted for \autoref{eq:CH4}, reduction of $\ce{CO_{(g)}}$ to $\ce{CH4_{(g)}}$ is severely kinetically inhibited over the lifetime of the solar nebula \citep{Lewis1980KINETICNEBULA}. The progress of solid-gas reactions is highly sensitive to grain size and temperature and they often follow a parabolic, diffusion-controlled kinetic. For example, the degree of enstatite formation from forsterite (\autoref{eq:orthopyroxene_condensation}) is expected to be very small during nebular timescales \citep{Imae1993AnGas}, whereas sulfidation of iron to troilite (\autoref{eq:FeS_reaction}) occurs on laboratory timescales of order ~$10^{2}$ years at 700~K and $10^{-3}$ bar \citep{Lauretta1996TheNebula}. Exploring non-equilibrium kinetics for condensates of reduced systems would therefore refine the mapping between stellar abundances and planetary building blocks, and represents an important direction for future work.

In addition, recent work has highlighted the role of grain-scale microphysical processes, showing that desorption of volatile species can depend sensitively on species-specific binding energies and grain surface properties. This can allow volatile species to remain adsorbed onto grain surfaces and be released at temperatures that differ from their single equilibrium condensation temperatures, thereby complicating the mapping between disk condensates and the bulk compositions of accreted planet(esimals) \citep{Tinacci2023TheoreticalDisks, Boitard-Crepeau2025WasFormation}. These non-equilibrium processes represent an important direction for future work.

The approach is also constrained by the scope of the FactSage thermochemical database \citep{Bale2016FactSage2010-2016}, which is limited to the species defined as pure phases and a finite set of solid solutions. For example, in the present calculations, iron silicides were treated as pure compounds; incorporating solid solution behaviour could further alter the condensation of elements soluble in these phases. 

\section{Conclusions}
In this study, we model condensation sequences across a range of stellar C/O ratios assumed to be representative of primordial protoplanetary disks, focusing on identifying the chemical reactions and redox pathways that govern solid formation. By explicitly including solid solutions and reduced species - silicides, carbides, nitrides, and sulfides - we capture a broader set of condensates relevant to chemically reduced environments than previously considered. Our model also tracks changes in oxygen fugacity, pressure effects, and how variations in model parameters ($T_0$ and $\sigma$) of stochastic accretion influence the bulk compositions of planet(esimal)s.

The key findings are as follows:
\begin{enumerate}
    \item We define three regimes based on formation of reduced condensates and their effect on disk chemistry: solar-like $(\ce{C/O}~$<$~0.7)$, transitional $(\ce{C/O} = 0.7-0.91)$, and reduced $(\ce{C/O}~$>$~0.92)$. Each regime exhibits distinct condensation sequences and $f\ce{O2}$ evolution with temperature.

    \item Solid solutions and $\ce{Fe_xSi_{(s)}}$ phases significantly influence the partitioning of Fe and Si between solid and gas and hence their condensation temperatures relative to solar-like sequences. Relative to earlier studies that neglected these phases, their inclusion at the same C/O ratio suppresses silicate- and SiC formation at high temperatures, especially in reduced systems.

    \item Both disk pressures and temperatures regulate the stability of volatiles and reduced condensates. Higher pressures suppress graphite and \ce{SiC} condensation due to the increasing stability of CH$_{4(g)}$ relative to CO$_{(g)}$, while sulfide formation is promoted.

    \item With increasing C/O, nebular condensation temperatures of metals (Fe, Ni) remain constant, whereas those of nominally refractory elements that condense into oxides (Al, Ca) decrease. The stability of TiC above C/O $>$0.92 causes a sharp increase in its condensation temperature. Both S and particularly C become more refractory (higher condensation temperature) as C/O increases, owing to the stability of (Fe,Mg,Ca)S and graphite.

    \item Owing to the non-negligible fractions of Si, O, C and S that condense over a wide temperature range without necessarily reaching 50~\% by mass, the use of $T_c^{50\%}$ for determining bulk planetary compositions is limited, particularly in high C/O ratio compositions.

    \item Planet(esimals) forming in high-C/O environments exhibit substantial variability in Fe/Si, Fe/Mg, Fe/O, and volatile content as a function of $T_0$. At higher $T_0$, compositions are Fe-enriched, consistent with elevated core mass fractions and potential Si partitioning into the core. At lower $T_0$, bulk compositions approach more Earth-like major-element ratios while retaining enhanced volatile inventories. This compositional diversity provides plausible building blocks for metal-enriched super-Mercury analogues and volatile-bearing rocky planets, underscoring the broader range of terrestrial outcomes possible in non-solar disk environments.

\end{enumerate}

By explicitly linking disk redox conditions to planetary bulk composition, this work provides a chemical framework that complements efforts to resolve the mass–radius degeneracy problem. We show that planets forming in low-f$\ce{O2}$ environments can acquire bulk compositions that diverge systematically from those implied by stellar abundances. Interpreting stellar compositions in a physically meaningful way, therefore, requires accounting for system-specifics.

\begin{acknowledgements}
This work was supported by the Swiss National Science Foundation (SNSF) through an Eccellenza Professorship (\#203668) and the Swiss State Secretariat for Education, Research and Innovation (SERI) under contract No. MB22.00033, a SERI-funded ERC Starting grant "2ATMO" to P.A.S. Parts of this work have been carried out within the framework of the National Centre of Competence in Research (NCCR) PlanetS supported by the SNSF under grant 51NF40\_205606.

\end{acknowledgements}

\bibliographystyle{aa.bst}
\bibliography{references}

\clearpage
\onecolumn
\begin{appendix} %First appendix

\section{Stellar abundance dataset} \label{sec:stellar abundance}

\begin{figure}[!h]
    \centering
    \includegraphics[width=0.8\linewidth]{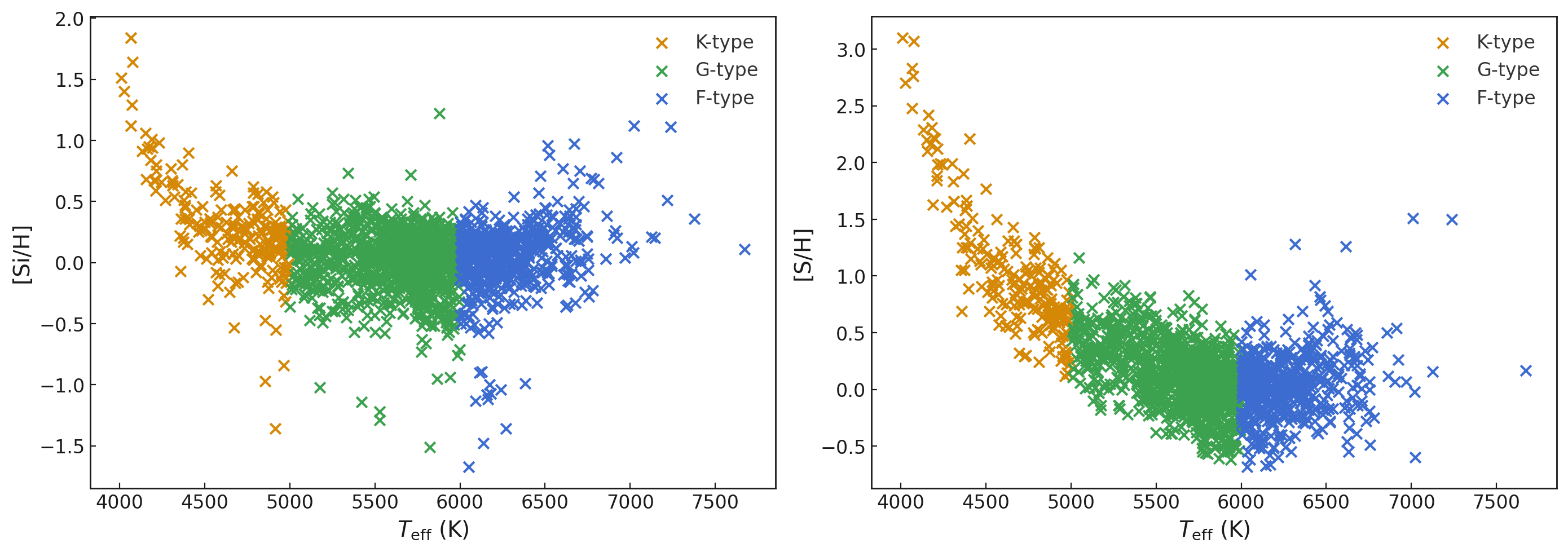}
    \caption{Elemental abundance vs. $T_{\text{eff}}$ (K) for FGK dwarf stars from \citet{Luck2017ABUNDANCESSUBGIANTS, Luck2018AbundancesDwarfs}}
    \label{fig:Teff_vs_abundance}
\end{figure}

Several abundance measurement studies (for eg. \citep{Luck2018AbundancesDwarfs, Luck2017ABUNDANCESSUBGIANTS}) which constitute databases like Hypatia Catalog \citep{Hinkel2014StellarCatalog} show systematic biases in the abundances of few elements, including [Si/H] and [S/H] such that they have higher abundance with lowe effective temperature (\autoref{fig:Teff_vs_abundance}). These biases may arise from strong line blending and inadequate $\log g$ values for stars at cooler temperatures \citep{Adibekyan2012ChemicalProgram}. 

In this study, we adopt elemental abundances of FGK dwarfs from the Portugal Group Dataset \citep{Adibekyan2012ChemicalProgram, DelgadoMena2017ChemicalProgram, CostaSilva2020ChemicalSample, DelgadoMena2021ChemicalProgram}, which correct these systematic trends for various elements. The corrections were applied by fitting a cubic polynomial and adding a constant term such that the correction is zero at solar temperature \citep{Adibekyan2012ChemicalProgram}.

Additionally, recent studies have highlighted errors in the calculation of oxygen abundances in stellar photospheres derived from the forbidden oxygen line at 6300~\AA, which can lead to the overestimation of C/O ratios \citep{Nissen2013ThePlanets, Fortney2012ONABUNDANCES}. To address this, we use carbon abundances derived from the lines at 5052~\AA\ and 5390~\AA. For oxygen, we use abundances derived from the $O\,\textsc{i}$ line at 6158~\AA, which provides a more conservative estimate compared to the forbidden line at 6300~\AA\ \citep{DelgadoMena2021ChemicalProgram}. Furthermore, we adopt a conservative approach by modelling only systems with C/O $< 1$, ensuring that our analysis focuses on systems where oxygen dominates over carbon in the gas phase.

\begin{table}[!h]
\centering
\caption{Elemental abundances (number fractions, mol \%) for PGD stellar sample used in this study.}
\label{tab:Stellar_data}
\resizebox{\textwidth}{!}{%
\begin{tabular}{lccccccccccccccccc}
\hline\hline
Star & N(H) & N(He) & N(Fe) & N(Na) & N(Mg) & N(Al) & N(Si) & N(Ca) & N(Ni) & N(S) & N(C) & N(O) & N(Ti) & N(Cr) & N(N) & C/O & [Fe/H] \\
\hline
HD24633   & 91.01 & 8.89 & 0.0039 & 0.0002 & 0.0050 & 0.0003 & 0.0032 & 0.0002 & 0.0001 & 0.0011 & 0.0359 & 0.0374 & $1.15\times10^{-5}$ & $4.31\times10^{-5}$ & 0.0048 & 0.959 & -0.04 \\
HD68607   & 91.02 & 8.89 & 0.0050 & 0.0002 & 0.0044 & 0.0003 & 0.0040 & 0.0003 & 0.0002 & 0.0014 & 0.0332 & 0.0355 & $1.08\times10^{-5}$ & $5.46\times10^{-5}$ & 0.0042 & 0.935 & 0.07 \\
HD90722   & 90.95 & 8.89 & 0.0087 & 0.0006 & 0.0074 & 0.0006 & 0.0071 & 0.0004 & 0.0004 & 0.0026 & 0.0553 & 0.0606 & $1.75\times10^{-5}$ & $8.71\times10^{-5}$ & 0.0135 & 0.912 & 0.31 \\
HD94151   & 91.03 & 8.90 & 0.0047 & 0.0002 & 0.0040 & 0.0003 & 0.0036 & 0.0002 & 0.0002 & 0.0013 & 0.0290 & 0.0322 & $9.73\times10^{-6}$ & $4.80\times10^{-5}$ & 0.0033 & 0.899 & 0.04 \\
HD13060   & 91.04 & 8.90 & 0.0045 & 0.0002 & 0.0039 & 0.0003 & 0.0034 & 0.0002 & 0.0002 & 0.0010 & 0.0239 & 0.0269 & $1.01\times10^{-5}$ & $4.86\times10^{-5}$ & 0.0020 & 0.887 & 0.02 \\
HD66340   & 91.02 & 8.90 & 0.0046 & 0.0002 & 0.0040 & 0.0003 & 0.0035 & 0.0002 & 0.0002 & 0.0012 & 0.0303 & 0.0342 & $1.00\times10^{-5}$ & $4.88\times10^{-5}$ & 0.0038 & 0.885 & 0.03 \\
HD72673   & 91.06 & 8.90 & 0.0017 & 0.0001 & 0.0017 & 0.0001 & 0.0015 & 0.0001 & 0.0001 & 0.0004 & 0.0154 & 0.0181 & $4.33\times10^{-6}$ & $1.81\times10^{-5}$ & 0.0005 & 0.849 & -0.41 \\
HD10895   & 91.07 & 8.90 & 0.0023 & 0.0001 & 0.0019 & 0.0001 & 0.0017 & 0.0001 & 0.0001 & 0.0007 & 0.0122 & 0.0145 & $5.33\times10^{-6}$ & $2.38\times10^{-5}$ & 0.0002 & 0.841 & -0.27 \\
HD7199    & 90.95 & 8.89 & 0.0081 & 0.0006 & 0.0075 & 0.0006 & 0.0074 & 0.0004 & 0.0004 & 0.0025 & 0.0522 & 0.0634 & $1.81\times10^{-5}$ & $8.99\times10^{-5}$ & 0.0146 & 0.824 & 0.28 \\
HD101367  & 90.97 & 8.89 & 0.0083 & 0.0006 & 0.0069 & 0.0006 & 0.0067 & 0.0004 & 0.0004 & 0.0022 & 0.0465 & 0.0567 & $1.82\times10^{-5}$ & $8.13\times10^{-5}$ & 0.0120 & 0.820 & 0.29 \\
HD202605  & 91.00 & 8.89 & 0.0064 & 0.0004 & 0.0049 & 0.0004 & 0.0048 & 0.0003 & 0.0003 & 0.0019 & 0.0344 & 0.0423 & $1.43\times10^{-5}$ & $6.92\times10^{-5}$ & 0.0064 & 0.813 & 0.18 \\
HD30306   & 90.99 & 8.89 & 0.0063 & 0.0003 & 0.0059 & 0.0004 & 0.0051 & 0.0003 & 0.0003 & 0.0016 & 0.0369 & 0.0466 & $1.39\times10^{-5}$ & $5.93\times10^{-5}$ & 0.0080 & 0.793 & 0.17 \\
HD18386   & 90.99 & 8.89 & 0.0059 & 0.0004 & 0.0048 & 0.0004 & 0.0055 & 0.0003 & 0.0003 & 0.0020 & 0.0394 & 0.0501 & $1.20\times10^{-5}$ & $6.22\times10^{-5}$ & 0.0094 & 0.787 & 0.14 \\
HD71835   & 91.03 & 8.90 & 0.0039 & 0.0002 & 0.0033 & 0.0002 & 0.0031 & 0.0002 & 0.0002 & 0.0011 & 0.0248 & 0.0324 & $7.84\times10^{-6}$ & $3.81\times10^{-5}$ & 0.0033 & 0.766 & -0.04 \\
HD48611   & 91.06 & 8.90 & 0.0019 & 0.0001 & 0.0020 & 0.0001 & 0.0016 & 0.0001 & 0.0001 & 0.0006 & 0.0150 & 0.0202 & $4.63\times10^{-6}$ & $1.93\times10^{-5}$ & 0.0008 & 0.741 & -0.36 \\
HD138549  & 91.02 & 8.89 & 0.0043 & 0.0002 & 0.0037 & 0.0003 & 0.0034 & 0.0002 & 0.0002 & 0.0013 & 0.0281 & 0.0386 & $9.31\times10^{-6}$ & $3.92\times10^{-5}$ & 0.0052 & 0.726 & 0.00 \\
HD100777  & 90.95 & 8.89 & 0.0076 & 0.0005 & 0.0067 & 0.0005 & 0.0070 & 0.0003 & 0.0003 & 0.0020 & 0.0476 & 0.0679 & $1.62\times10^{-5}$ & $6.70\times10^{-5}$ & 0.0164 & 0.701 & 0.25 \\
HD100289  & 91.01 & 8.89 & 0.0046 & 0.0003 & 0.0040 & 0.0003 & 0.0036 & 0.0002 & 0.0002 & 0.0016 & 0.0288 & 0.0412 & $1.05\times10^{-5}$ & $4.27\times10^{-5}$ & 0.0061 & 0.698 & 0.03 \\
HD24085   & 91.00 & 8.89 & 0.0063 & 0.0003 & 0.0051 & 0.0004 & 0.0049 & 0.0003 & 0.0003 & 0.0016 & 0.0330 & 0.0483 & $1.44\times10^{-5}$ & $6.93\times10^{-5}$ & 0.0087 & 0.682 & 0.17 \\
HD207583  & 91.03 & 8.90 & 0.0044 & 0.0002 & 0.0033 & 0.0002 & 0.0032 & 0.0002 & 0.0002 & 0.0011 & 0.0222 & 0.0338 & $9.08\times10^{-6}$ & $3.89\times10^{-5}$ & 0.0037 & 0.658 & 0.01 \\
HD131664  & 90.96 & 8.89 & 0.0087 & 0.0005 & 0.0069 & 0.0006 & 0.0067 & 0.0004 & 0.0004 & 0.0022 & 0.0432 & 0.0660 & $1.87\times10^{-5}$ & $8.31\times10^{-5}$ & 0.0157 & 0.655 & 0.31 \\
Lodders03 & 91.12 & 8.78 & 0.0032 & 0.0002 & 0.0038 & 0.0003 & 0.0037 & 0.0002 & 0.0002 & 0.0017 & 0.0263 & 0.0524 & $9.11\times10^{-6}$ & $4.78\times10^{-5}$ & 0.0102 & 0.501 & 0.00 \\
\hline
\end{tabular}%
}
\tablefoot{Values are normalized such that the sum of all species equals 100.}
\end{table}

\section{Chemical species included from FactSage}
\label{sec:Factsage_species}
\begin{table}
\centering
\caption{Ideal Gas species used in Equilibrium Calculations.}
\label{tab:gas_species}
\begin{tabular}{cccccc}
\toprule
\ce{H} & \ce{H2} & \ce{He} & \ce{C} & \ce{C2} & \ce{C3} \\
\ce{C4} & \ce{C5} & \ce{CH} & \ce{CH2} & \ce{CH3} & \ce{CH4} \\
\ce{C2H} & \ce{C2H2} & \ce{C2H3} & \ce{C2H4} & \ce{C2H5} & \ce{C2H6} \\
\ce{N} & \ce{N2} & \ce{N3} & \ce{NH} & \ce{NH2} & \ce{NH3} \\
\ce{HNNH} & \ce{N2H4} & \ce{CN} & \ce{C2N} & \ce{CNN} & \ce{CNN(g2)} \\
\ce{(CN)2} & \ce{C4N2} & \ce{HCN} & \ce{CH3NH2} & \ce{HCCN} & \ce{CH3NC} \\
\ce{C2H5N} & \ce{(CH3)2NH} & \ce{CH3N2H3} & \ce{O} & \ce{O2} & \ce{O3} \\
\ce{OH} & \ce{H2O} & \ce{HOO} & \ce{HOOH} & \ce{CO} & \ce{C2O} \\
\ce{CO2} & \ce{C3O2} & \ce{HCO} & \ce{H2CO} & \ce{CH3O} & \ce{CH3O(g2)} \\
\ce{CH3OH} & \ce{CH2CO} & \ce{C2H4O} & \ce{C2H4O(g2)} & \ce{CH3CH2OH} & \ce{CH3CH2OH(g2)} \\
\ce{COOH} & \ce{HCOOH} & \ce{CH3COOH} & \ce{NO} & \ce{N2O} & \ce{NO2} \\
\ce{NO3} & \ce{N2O3} & \ce{N2O4} & \ce{N2O5} & \ce{HNO} & \ce{N2H5OH} \\
\ce{HONO} & \ce{HONO(g2)} & \ce{HONO2} & \ce{NCO} & \ce{HNCO} & \ce{CH3NO2} \\
\ce{CH3CH2ONO2} & \ce{Na} & \ce{Na2} & \ce{NaH} & \ce{NaCN} & \ce{(NaCN)2} \\
\ce{NaO} & \ce{NaOH} & \ce{(NaOH)2} & \ce{Mg} & \ce{Mg2} & \ce{MgH} \\
\ce{MgN} & \ce{MgO} & \ce{MgOH} & \ce{Mg(OH)2} & \ce{Al} & \ce{Al2} \\
\ce{AlH} & \ce{AlC} & \ce{AlC2} & \ce{Al2C2} & \ce{AlN} & \ce{AlO} \\
\ce{AlO2} & \ce{Al2O} & \ce{Al2O2} & \ce{Al2O3} & \ce{AlOH} & \ce{AlOH(g2)} \\
\ce{OAlOH} & \ce{Si} & \ce{Si2} & \ce{Si3} & \ce{SiH} & \ce{SiH4} \\
\ce{Si2H6} & \ce{SiC} & \ce{SiC2} & \ce{Si2C} & \ce{SiN} & \ce{Si2N} \\
\ce{SiO} & \ce{SiO2} & \ce{S} & \ce{S2} & \ce{S3} & \ce{S4} \\
\ce{S5} & \ce{S6} & \ce{S7} & \ce{S8} & \ce{HS} & \ce{H2S} \\
\ce{H2S2} & \ce{CS} & \ce{CS2} & \ce{CH3SH} & \ce{C2H4S} & \ce{(CH3)2S} \\
\ce{(CH3)2S(g2)} & \ce{CH3SSCH3} & \ce{NS} & \ce{HCNS} & \ce{CH3NCS} & \ce{SO} \\
\ce{SO2} & \ce{SO3} & \ce{SSO} & \ce{H2SO4} & \ce{COS} & \ce{C2H4OS} \\
\ce{(CH3)2SO} & \ce{(CH3)2SO2} & \ce{Na2SO4} & \ce{MgS} & \ce{AlS} & \ce{Al2S} \\
\ce{Al2S2} & \ce{SiS} & \ce{SiS2} & \ce{Ca} & \ce{Ca2} & \ce{CaH} \\
\ce{CaO} & \ce{CaOH} & \ce{Ca(OH)2} & \ce{CaS} & \ce{Ti} & \ce{TiO} \\
\ce{TiO2} & \ce{TiS} & \ce{Cr} & \ce{CrN} & \ce{CrO} & \ce{CrO2} \\
\ce{CrO3} & \ce{CrOH} & \ce{CrOOH} & \ce{Cr(OH)2} & \ce{CrO2OH} & \ce{CrO(OH)2} \\
\ce{Cr(OH)3} & \ce{CrO2(OH)2} & \ce{CrO(OH)3} & \ce{Cr(OH)4} & \ce{CrO(OH)4} & \ce{Cr(OH)5} \\
\ce{Cr(OH)6} & \ce{CrS} & \ce{Fe} & \ce{FeO} & \ce{Fe(OH)2} & \ce{Fe(CO)5} \\
\ce{FeS} & \ce{Ni} & \ce{NiH} & \ce{NiO} & \ce{Ni(OH)2} & \ce{Ni(CO)4} \\
\ce{NiS} & & & & & \\
\bottomrule
\end{tabular}
\end{table}

% Switch to single-column mode for the table

\begin{longtable}{p{3cm} p{12cm}}
\caption{Solid Solutions used in Equilibrium Calculations.} \label{tab:solid_soln} \\
\toprule
\textbf{FactSage Code} & \textbf{Species} \\
\midrule
\endfirsthead
\toprule
\textbf{FactSage Code} & \textbf{Species} \\
\midrule
\endhead
\midrule
\endfoot
\bottomrule
\endlastfoot
\textbf{FTOxCN-AlC\_} & \ce{Al4C3}, \ce{Si4C3[4+]}, \ce{Va4C3[12-]} \\
\midrule
\textbf{FTOxCN-CMFS} & \ce{CaS}, \ce{MgS}, \ce{FeS} \\
\midrule
\textbf{FToxid-SPINA} & \ce{Fe3O4}, \ce{Fe3O4[1-]}, \ce{Fe3O4[1+]}, \ce{Fe3O4[2-]}, \ce{Fe1O4[5-]}, \ce{Fe1O4[6-]}, \ce{Fe1Al2O4}, \ce{Al3O4[1+]}, \ce{Al1Fe2O4[1-]}, \ce{Al1O4[5-]}, \ce{Fe1Al2O4[1+]}, \ce{Al1Fe2O4[1+]}, \ce{Mg1Al2O4}, \ce{Al1Mg2O4[1-]}, \ce{Mg3O4[2-]}, \ce{Mg1O4[6-]}, \ce{Mg1Fe2O4}, \ce{Fe1Mg2O4[1-]}, \ce{Fe1Mg2O4[2-]}, \ce{Mg1Fe2O4[2-]}, \ce{Mg1Cr2O4}, \ce{Fe1Cr2O4}, \ce{Cr1Cr2O4[1+]}, \ce{Cr1Mg2O4[1-]}, \ce{Cr1Fe2O4[1-]}, \ce{Fe1Cr2O4[1+]}, \ce{Cr1Fe2O4[1+]}, \ce{Al1Cr2O4[1+]}, \ce{Cr1Al2O4[1+]}, \ce{Cr1O4[5-]}, \ce{Cr1Cr2O4}, \ce{Cr1Fe2O4[2-]}, \ce{Cr1Mg2O4[2-]}, \ce{Cr1Fe2O4}, \ce{Cr1Al2O4}, \ce{Cr1O4[6-]}, \ce{Ni1Al2O4}, \ce{Al1Ni2O4[1-]}, \ce{Ni3O4[2-]}, \ce{Ni1O4[6-]}, \ce{Ni1Fe2O4}, \ce{Fe1Ni2O4[1-]}, \ce{Fe1Ni2O4[2-]}, \ce{Ni1Fe2O4[2-]}, \ce{Ni1Cr2O4}, \ce{Cr1Ni2O4[1-]}, \ce{Mg1Ni2O4[2-]}, \ce{Ni1Mg2O4[2-]}, \ce{Cr1Ni2O4[2-]} \\
\midrule
\textbf{FToxid-cPyr} & \ce{CaMgSi2O6}, \ce{CaFeSi2O6}, \ce{FeFeSi2O6}, \ce{MgMgSi2O6}, \ce{CaAl2SiO6}, \ce{CaFe3+AlSiO6}, \ce{MgFeSi2O6}, \ce{FeMgSi2O6}, \ce{CaAlFe3+SiO6}, \ce{CaFe3+Fe3+SiO6}, \ce{FeAlAlSiO6}, \ce{MgAlAlSiO6}, \ce{FeFe3+Fe3+SiO6}, \ce{FeFe3+AlSiO6}, \ce{FeAlFe3+SiO6}, \ce{MgFe3+Fe3+SiO6}, \ce{MgFe3+AlSiO6}, \ce{MgAlFe3+SiO6}, \ce{FeFe3+Si2O6[+]}, \ce{CaFe3+Si2O6[+]}, \ce{MgFe3+Si2O6[+]} \\
\midrule
\textbf{FToxid-oPyr} & \textit{Identical to FToxid-cPyr components} \\
\midrule
\textbf{FToxid-WOLL} & \ce{MgSiO3}, \ce{FeSiO3}, \ce{CaSiO3} \\
\midrule
\textbf{FToxid-Oliv} & \ce{Mg1Mg1Si1O4}, \ce{Fe1Fe1Si1O4}, \ce{Mg1Fe1Si1O4}, \ce{Fe1Mg1Si1O4}, \ce{Ca1Ca1Si1O4}, \ce{Ca1Fe1Si1O4}, \ce{Fe1Ca1Si1O4}, \ce{Ca1Mg1Si1O4}, \ce{Mg1Ca1Si1O4}, \ce{Ca1Ni1Si1O4}, \ce{Ni1Ca1Si1O4}, \ce{Mg1Ni1Si1O4}, \ce{Ni1Mg1Si1O4}, \ce{Fe1Ni1Si1O4}, \ce{Ni1Fe1Si1O4}, \ce{Ni1Ni1Si1O4} \\
\midrule
\textbf{FToxid-Cord} & \ce{Al4Fe2Si5O18}, \ce{Al4Mg2Si5O18} \\
\midrule
\textbf{FToxid-CAFS} & \ce{Ca2Fe8Si1O16}, \ce{Ca2Al8Si1O16} \\
\midrule
\textbf{FToxid-CAF6} & \ce{Ca1Al12O19}, \ce{Ca1Fe12O19} \\
\midrule
\textbf{FToxid-CAF3} & \ce{Ca1Al6O10}, \ce{Ca1Fe6O10} \\
\midrule
\textbf{FToxid-CAF2} & \ce{Ca1Al4O7}, \ce{Ca1Fe4O7} \\
\midrule
\textbf{FToxid-CAF1} & \ce{Ca1Al2O4}, \ce{Ca1Fe2O4} \\
\midrule
\textbf{FToxid-C2AF} & \ce{Ca2Al2O5}, \ce{Ca2Fe2O5} \\
\midrule
\textbf{FToxid-C3AF} & \ce{Ca3Al2O6}, \ce{Ca3Fe2O6} \\
\midrule
\textbf{FToxid-CORU} & \ce{Al2O3}, \ce{Cr2O3}, \ce{Fe2O3}, \ce{Ti2O3}, \ce{NiO} \\
\midrule
\textbf{FToxid-GARN} & \ce{Ca3Cr2Si3O12}, \ce{Ca3Al2Si3O12} \\
\midrule
\textbf{FToxid-CaSp} & \ce{CaCr2O4}, \ce{CaFe2O4} \\
\midrule
\textbf{FToxid-Carn} & \ce{NaAlSiO4}, \ce{Si2O4}, \ce{CaAl2O4}, \ce{CaAl2Si2O8}, \ce{NaFeSiO4} \\
\midrule
\textbf{FToxid-Neph} & \textit{Identical to FToxid-Carn components} \\
\midrule
\textbf{FToxid-Feld} & \ce{NaAlSi3O8}, \ce{CaAl2Si2O8}, \ce{NaFeSi3O8} \\
\midrule
\textbf{FToxid-TiO2} & \ce{Ti2O3}, \ce{TiO2} \\
\midrule
\textbf{FToxid-ILME} & \ce{MgTiO3[-]}, \ce{FeTiO3[-]}, \ce{Ti2O3[+]}, \ce{MgTiO3}, \ce{FeTiO3}, \ce{Ti2O3} \\
\midrule
\textbf{FToxid-TiSp} & \ce{Mg3O4[2-]}, \ce{FeMg2O4[2-]}, \ce{MgFe2O4[2-]}, \ce{Fe3O4[2-]}, \ce{MgTi2O4}, \ce{FeTi2O4}, \ce{MgTi2O4[2+]}, \ce{FeTi2O4[2+]} \\
\midrule
\textbf{FToxid-CaTi} & \ce{Ca3Ti2O7}, \ce{Ca3Ti2O6} \\
\midrule
\textbf{FToxid-PERO} & \ce{Ca2Ti2O6}, \ce{Ca2Ti2O5} \\
\midrule
\textbf{FTsalt-oS24} & \ce{MgCrO4}, \ce{CaCrO4}, \ce{MgSO4}, \ce{CaSO4_anhydrite} \\
\midrule
\textbf{FTsulf-PYRR} & SVa, SFe, SNi, SCr \\
\midrule
\textbf{FTsulf-MeS2} & \ce{FeS2}, \ce{NiS2} \\
\midrule
\textbf{FTsulf-Pent} & \ce{Fe9S8}, \ce{Ni9S8} \\
\midrule
\textbf{FTsulf-BCCS} & Cr:Va, Fe:Va, Ni:Va, S:Va, O:Va \\
\midrule
\textbf{FTsulf-FCCS} & Cr:Va, Fe:Va, Ni:Va, S:Va, O:Va \\
\midrule
\textbf{Ftmisc-FeS} & Fe, FeS, FeO, MgS, TiS, \ce{Na2S} \\
\midrule
\textbf{Ftoxid-NCS0} & \ce{Na2Na2Ca3CaSi6O18}, \ce{Na2Na2Na6CaSi6O18}, \ce{Ca1Na2Ca3CaSi6O18}, \ce{Ca1Na2Na6CaSi6O18} \\
\midrule
\textbf{FTOxCN-HEXA} & \ce{Al5C3N}, \ce{Al4SiC4} \\
\midrule
\textbf{FTOxCN-AlON} & \ce{Al3N4[3-]}, \ce{Al1N4[9-]}, \ce{Al3O4[1+]}, \ce{Al1O4[5-]} \\
\midrule
\textbf{FTOxCN-A2OC} & \ce{Al2OC}, \ce{Al2N2} \\
\midrule
\textbf{FTOxCN-SiNO} & \ce{Si2N2O}, \ce{Al2O3} \\
\midrule
\textbf{FTsulf-M3S2} & Va2S, Fe2S, Ni2S \\
\midrule
\textbf{FTOxCN-Beta} & \ce{Si3N4}, \ce{Al3O3N} \\
\end{longtable}

\begin{table}[!ht]
\centering
\caption{Pure solid species used in Equilibrium Calculations.}
\label{tab:solid_species}
\begin{tabular}{cccc}
\toprule
\ce{Al2S3} & \ce{C(s)} & \ce{Ca(s)} & \ce{Ca2MgSi2O7(s)} \\
\ce{Cr(s)} & \ce{Cr2N(s)} & \ce{Cr3C2(s)} & \ce{Cr4C(s)} \\
\ce{CrN(s)} & \ce{Fe3C(s)} & \ce{Fe3Si(s)} & \ce{Fe5C2(s)} \\
\ce{FeCr2O4(s)} & \ce{FeCr2S4(s)} & \ce{FeSi(s)} & \ce{Mg(s)} \\
\ce{Mg4Al10Si2O23(s)} & \ce{(MgO)(Cr2O3)(s)} & \ce{Na(s)} & \ce{NaAlO2(s)} \\
\ce{S(s)} & \ce{Si(s)} & \ce{SiC(s)} & \ce{SiC(S2)(s)} \\
\ce{TiC(s)} & \ce{TiN(s)} & & \\
\bottomrule
\end{tabular}
\end{table}

%%%%%%%%%%%%%%%%%%%%%%%%%%%%%%%%%%%%%%%%%%%%%%%%%%%%%
\section{Supplementary figures} \label{sec:disk-chem-appendix}
The figures in this appendix support and extend specific results from Sects .~\ref {sec:disk_chem_composition}--\ref{sec:planet_composition}.

\subsection{Supplementary plots corresponding to Section \ref{sec:disk_chem_composition}}

\begin{figure}[!ht]
    \centering
    \includegraphics[width=1\linewidth]{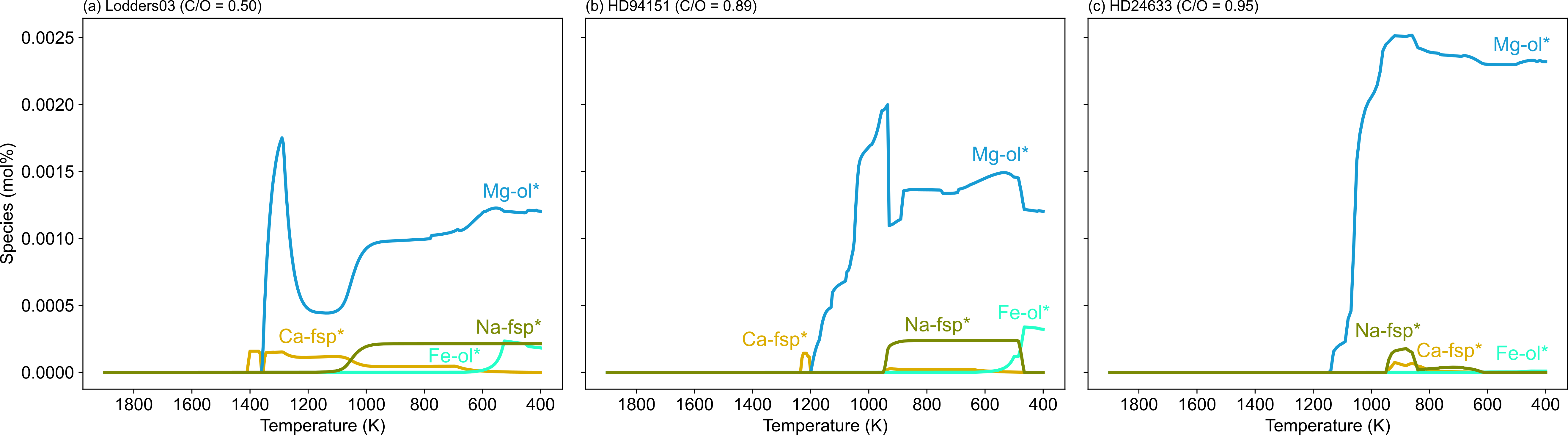}
    \caption{Solid Solution of olivine and feldspar at a disk pressure of $10^{-4}$ bar.}
    \label{fig:silicate_solidsoln_comp_linear}
\end{figure}

\begin{figure}[!ht]
    \centering
    \includegraphics[width=1\linewidth]{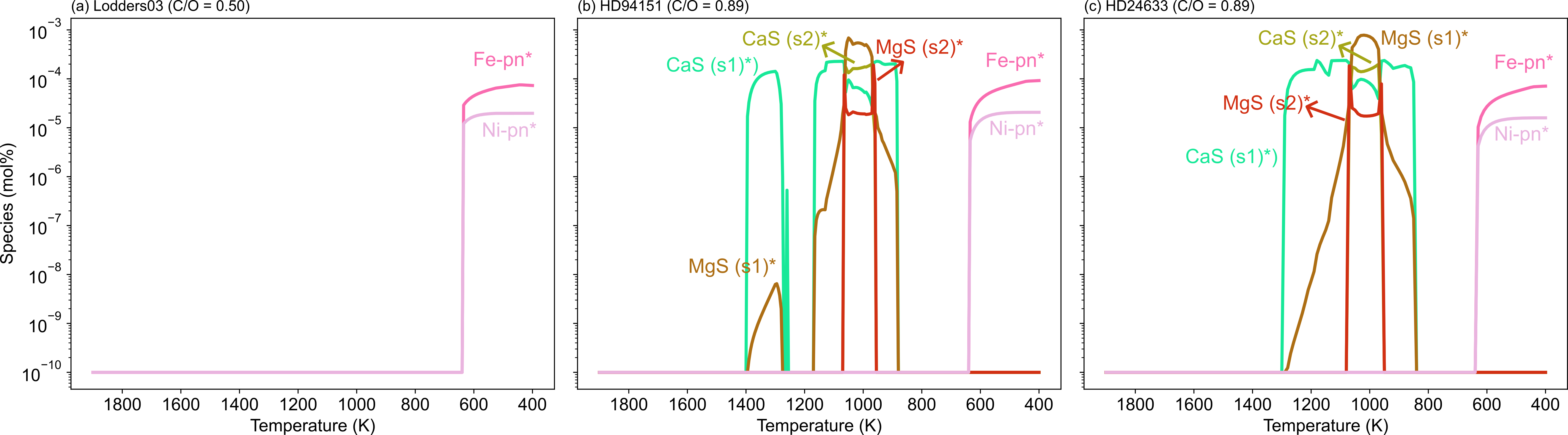}
    \caption{Solid Solution components of $\ce{(Ca,Mg)S_{(s)}}$ at a disk pressure of $10^{-4}$ bar}.
    \label{fig:sulfur_solidsoln_comp_linear}
\end{figure}

\begin{figure}[!ht]
    \centering
    \includegraphics[width=1\linewidth]{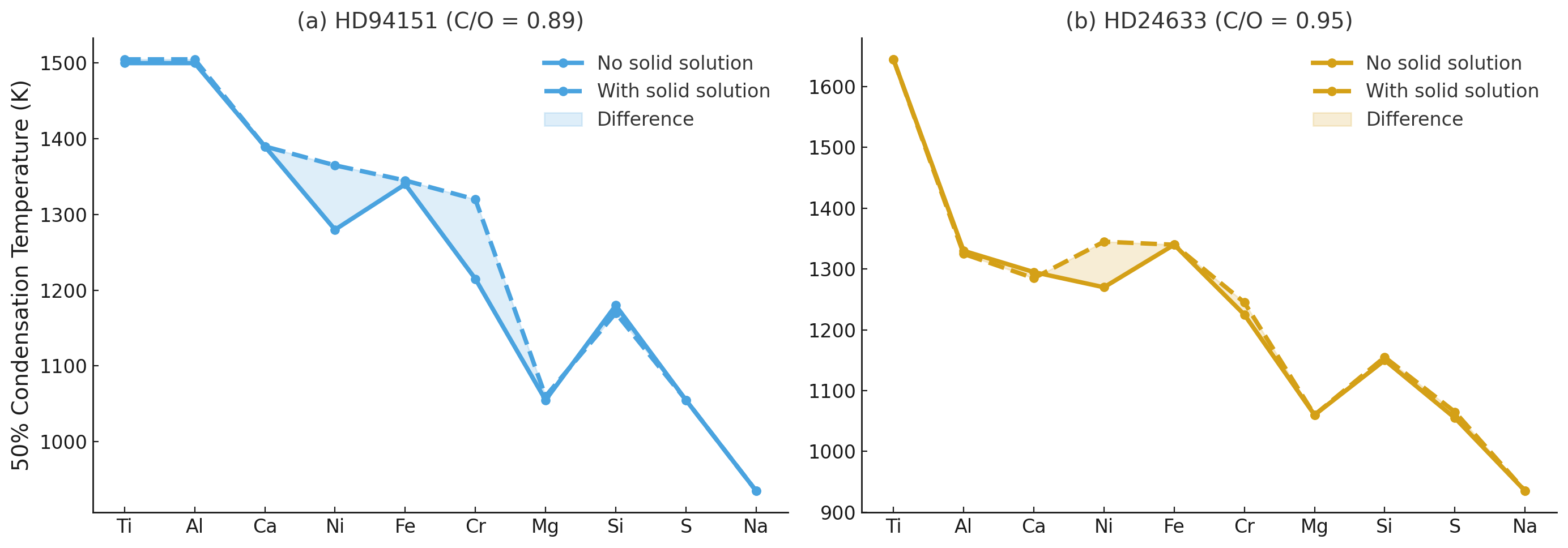}
    \caption{50\% condensation temperatures of systems HD94151 and HD24633 shown for condensation calculations with and without solid solutions.}
    \label{fig:condensation_solidsoln}
\end{figure}

\begin{figure}[!ht]
    \centering
    \includegraphics[width=1\linewidth]{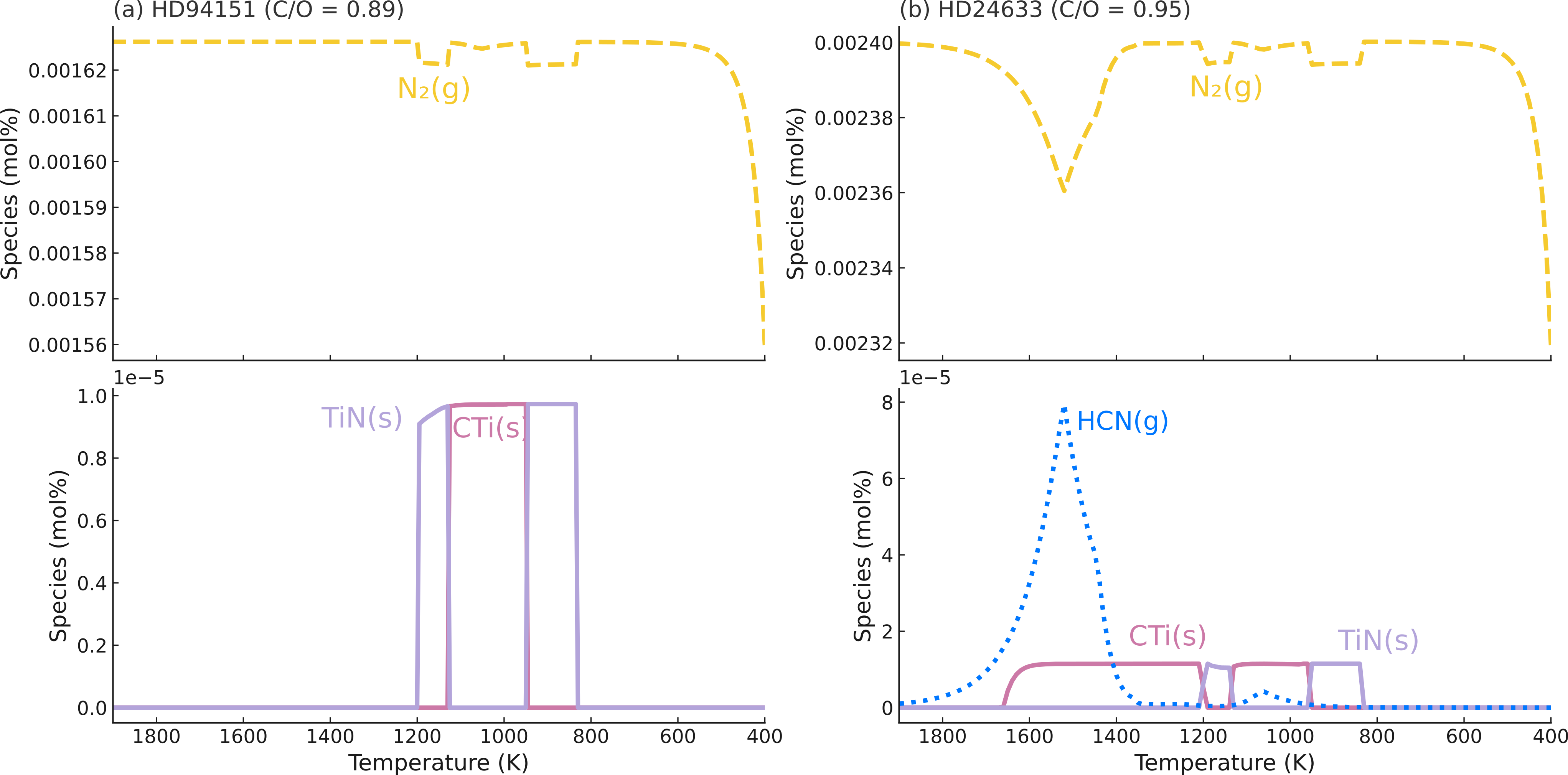}
    \caption{Comparison of nitrogen-bearing species in transitional (HD94151; C/O$=0.89$) and reduced (HD24633; C/O$=0.95$) environments at $10^{-4}$~bar. Upper panels show $\ce{N2_{(g)}}$; lower panels display condensed $\ce{TiN_{(s)}}$ and $\ce{TiC_{(s)}}$, with $\ce{HCN_{(g)}}$ additionally present in the reduced model. The opposing trends of $\ce{TiN_{(s)}}$ and $\ce{TiC_{(s)}}$ reflect the conversion between these condensates.}
    \label{fig:nitrogen_condensation_comparision}
\end{figure}

\begin{figure}[!ht]
    \centering
    \includegraphics[width=0.8\linewidth]{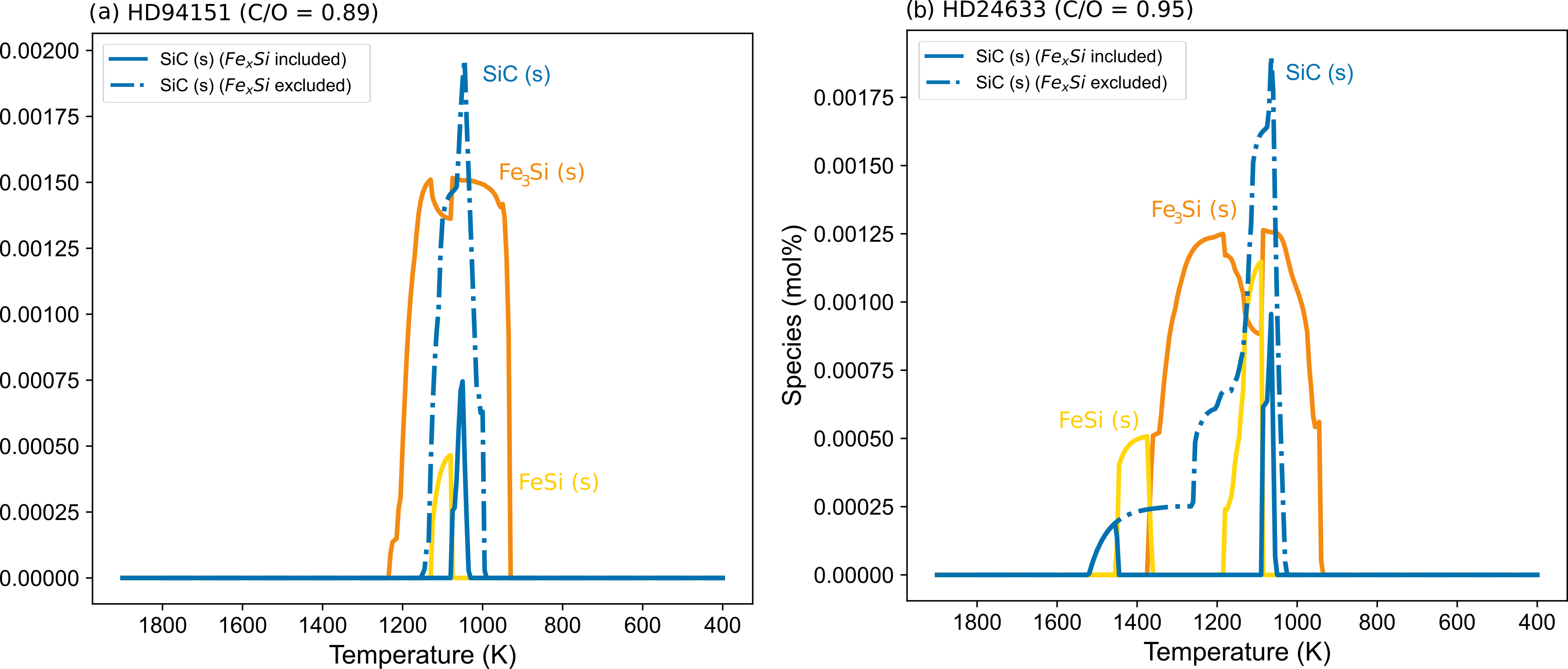}
    \caption{Condensation Curve for \ce{FeSi}, \ce{Fe3Si} and \ce{SiC} in systems HD24633 and HD94151 to compare stability of \ce{SiC} with and without iron silicides in condensation calculations.}
    \label{fig:FexSi_comparision}
\end{figure}

\FloatBarrier
\clearpage
\subsection{Supplementary plots corresponding to Section \ref{sec:disk_chem_pressure}}

\begin{figure}[!ht]
    \centering
    \includegraphics[width=1\linewidth]{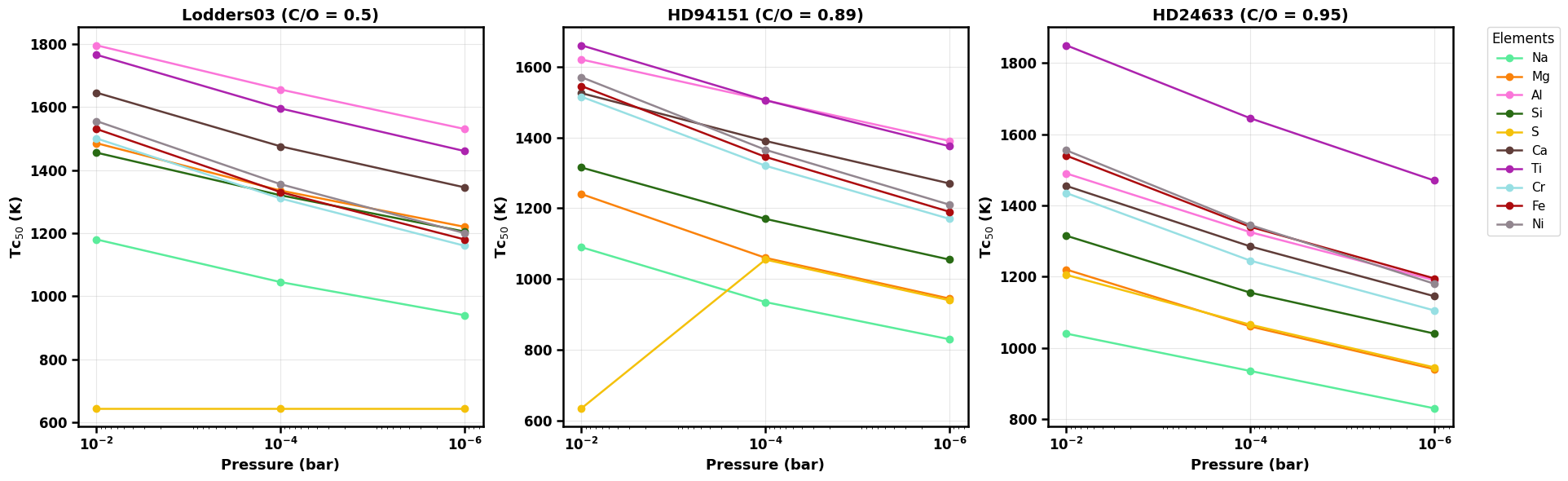}
    \caption{Condensation temperatures ($T_c^{50\%}$) for solar-like (Lodders03; C/O$= 0.50$), transitional (HD94151; C/O$= 0.89$), and reduced (HD24633; C/O$= 0.95$) systems at $10^{-2}$, $10^{-4}$ and $10^{-6}$ bar disk pressures.}
    \label{fig:tc50_vs_pressure}
\end{figure}

\begin{figure}[!ht]
    \centering
    \includegraphics[width=1\linewidth]{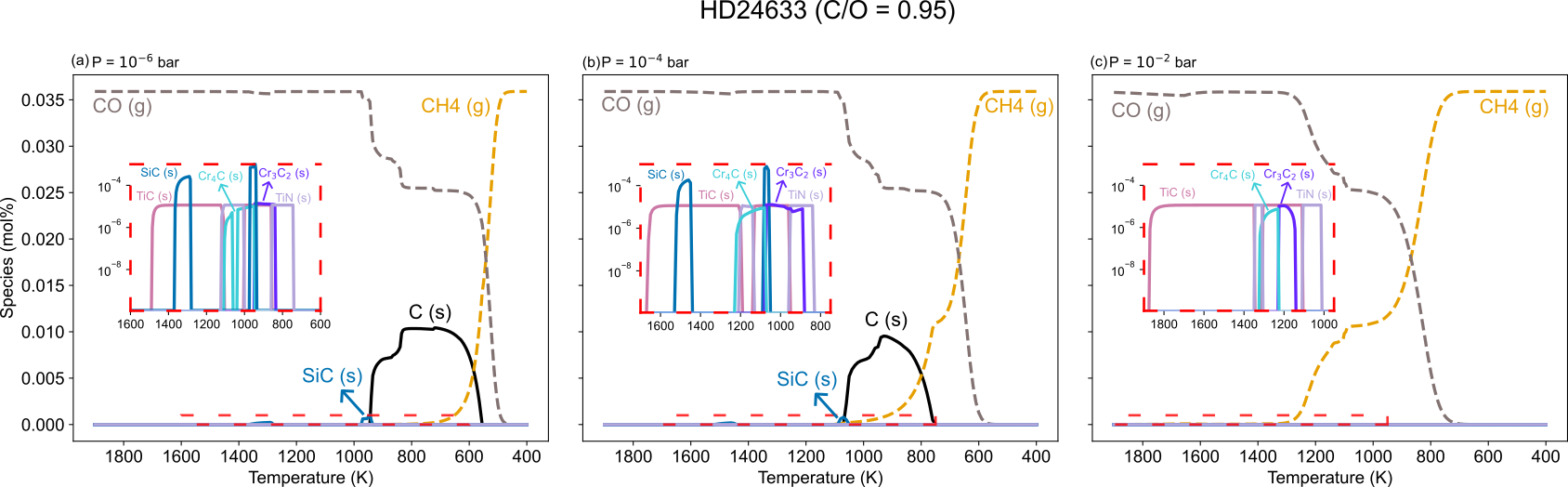}
    \caption{Stability of carbon-bearing phases in system HD24633 (C/O = 0.95) at different pressures.}
    \label{fig:HD24633_pres_carbon}
\end{figure}

\begin{figure}[!ht]
    \centering
    \includegraphics[width=1\linewidth]{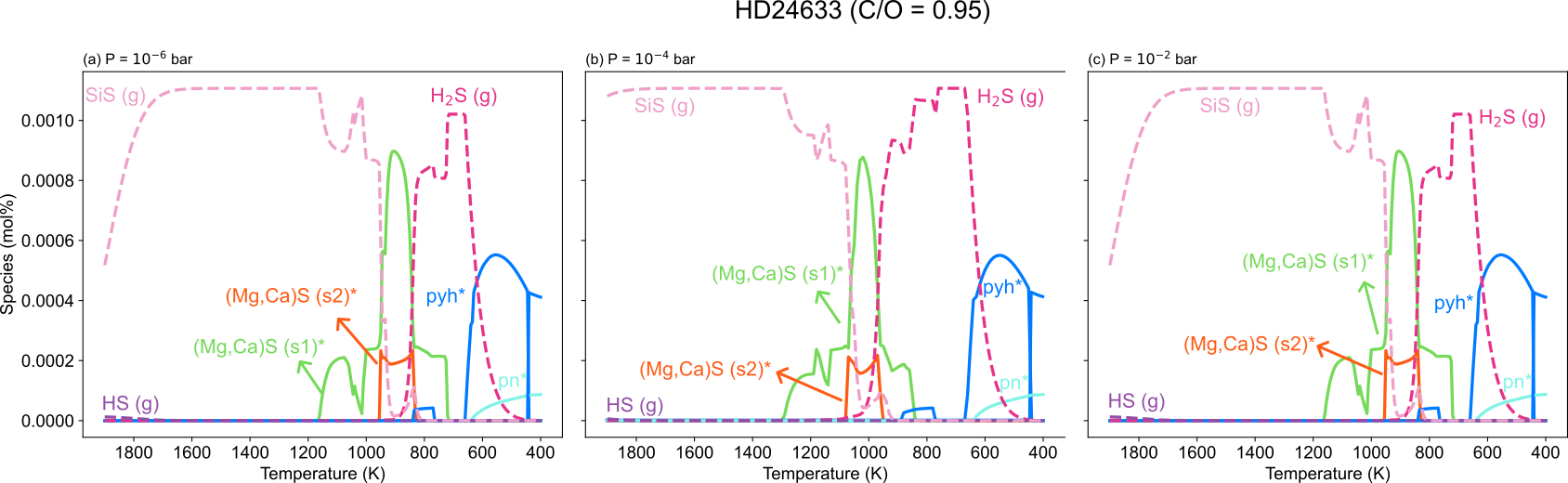}
    \caption{Stability of sulfur-bearing phases in system HD24633 (C/O = 0.95) at different pressures.}
    \label{fig:HD24633_pres_sulfur}
\end{figure}

\FloatBarrier
\clearpage
\subsection{Supplementary plots corresponding to Section \ref{sec:planet_composition}}

\begin{figure}[!ht]
    \centering
    \includegraphics[width=1\linewidth]{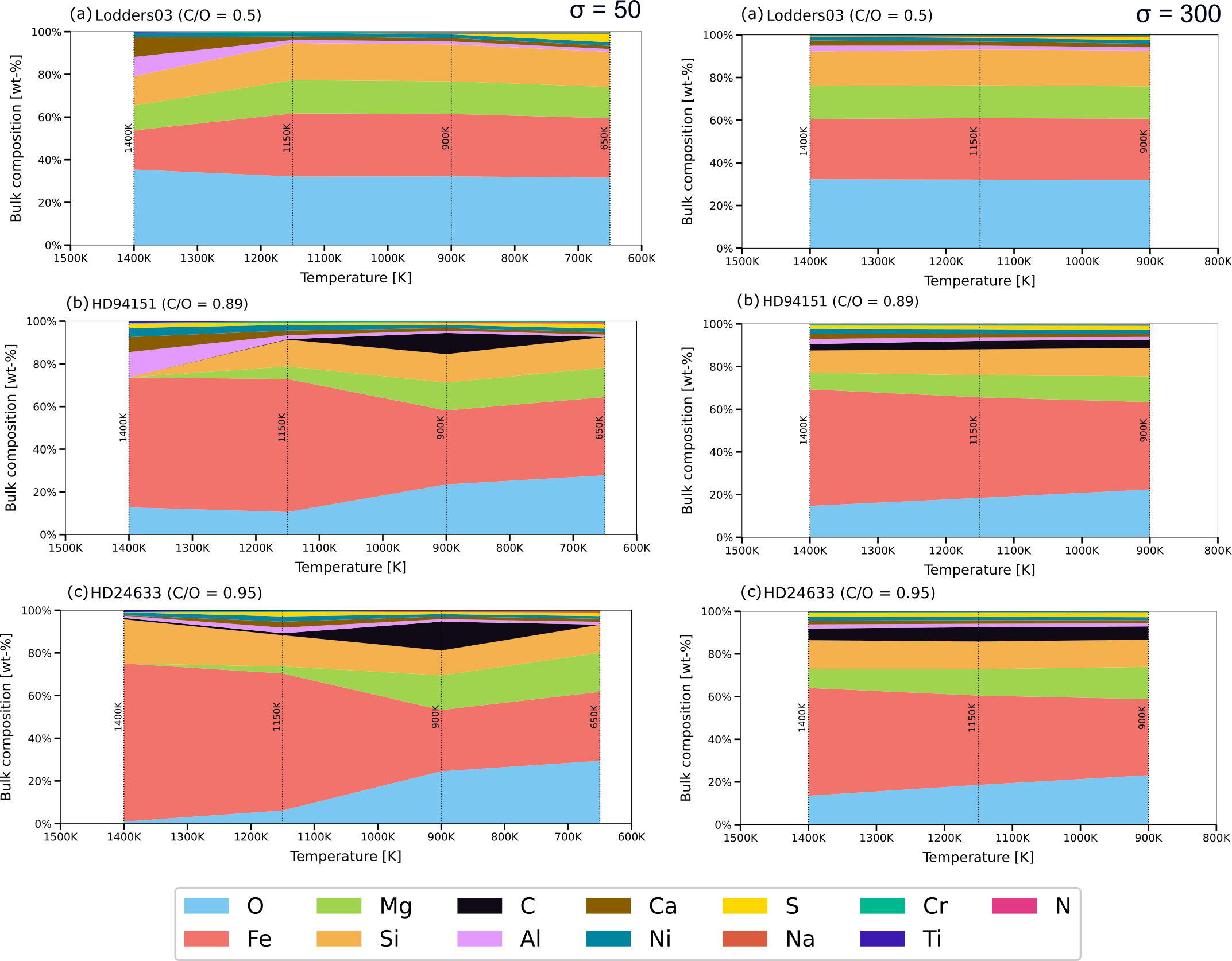}
    \caption{Same as \autoref{fig:t0_planet} but for $\sigma = 50$ (left column) and $\sigma = 300$ (right column). Vertical dashed lines indicate the $T_0$ values adopted in the accretion calculations.}
    \label{fig:sigma_comparision}
\end{figure}

\begin{figure}[!ht]
    \centering
    \includegraphics[width=0.7\linewidth]{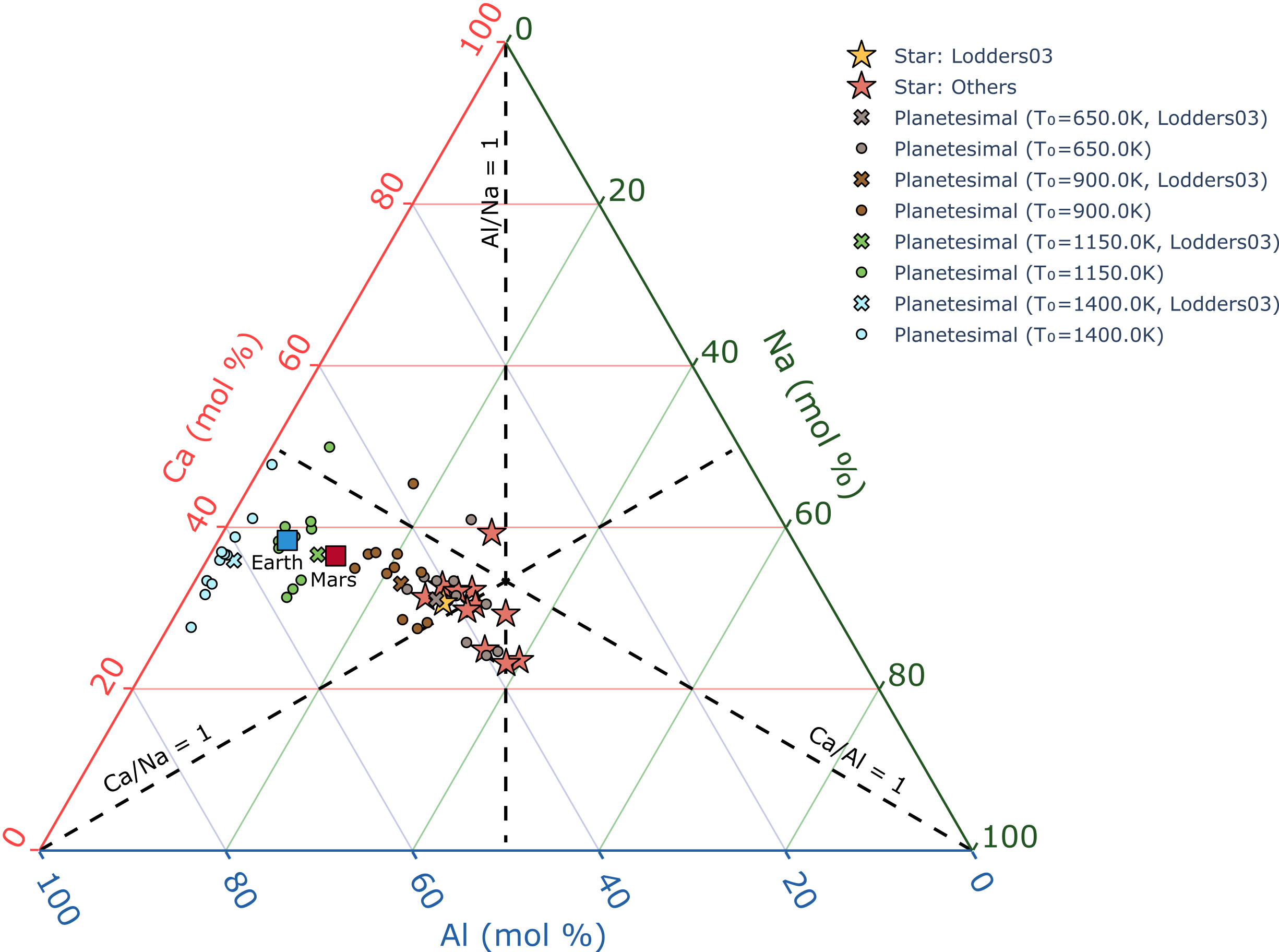}
    \caption{Ternary diagram Ca-Al-Na (mol \%) displaying high C/O stars and modeled planet(esimal)s. Sun \citep{Lodders2003SolarElements}, Earth \citep{Fischer2020TheCore} and Mars \citep{Yoshizaki2020TheMars} are shown for reference.}
    \label{fig:Ca-Al-Na_ternary}
\end{figure}

\autoref{fig:Ca-Al-Na_ternary} illustrates that the stars exhibit a wide variation in the Ca/Al ratios, which is proportionally reflected in the resultant planet(esimals), with $(\text{Ca/Al})_\text{star}$ being approximately proportional to $(\text{Ca/Al})_\text{planet}$. On average, the Ca/Al ratio is $0.79 \pm 0.16$ for stars and $0.72 \pm 0.17$ for planet(esimals). Within a given system, the Ca/Al ratio remains relatively constant but increases at lower $T_0$. Notably, the Ca/Al ratio of planet(esimals) aligns with the stellar ratio at formation temperatures between $650~\text{K}$ and $900~\text{K}$. Sodium, as a moderately volatile element, is significantly depleted at higher $T_0$ values. This trend is consistent for both the solar-like systems and high C/O ratio systems.

\begin{figure}[ht!]
    \centering
    \includegraphics[width=0.8\linewidth]{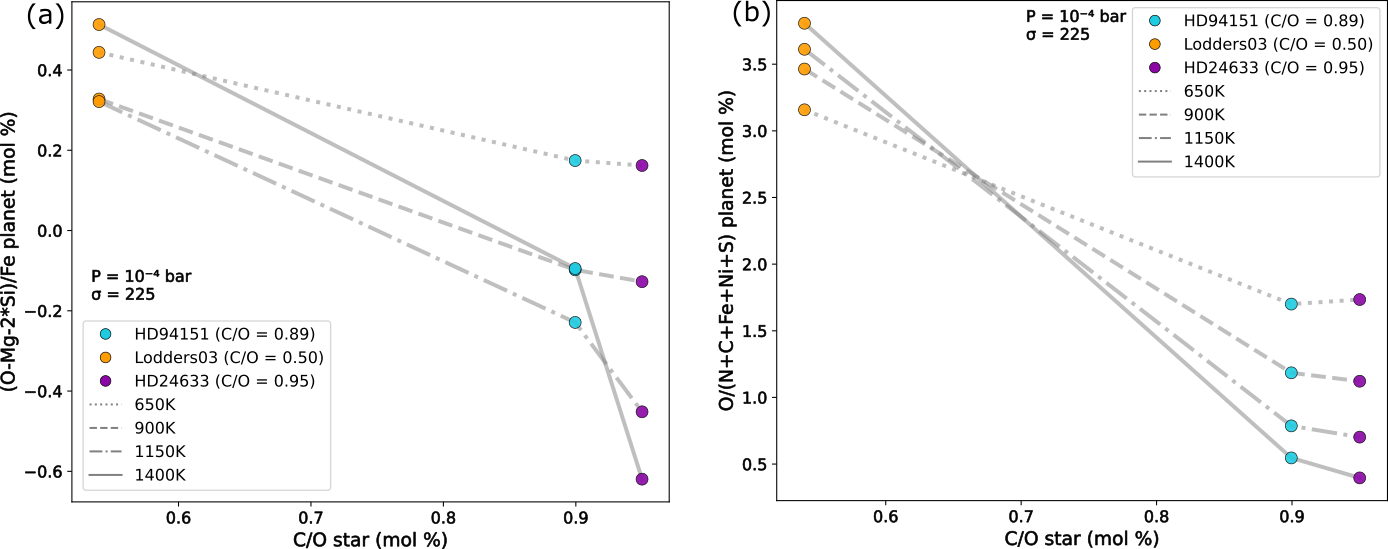}
    \caption{(a) Factor $(O-Mg-2Si)/Fe$ for planet(esimals) (mol \%) against C/O ratio of the star for disk pressure of $10^{-4}$ bar and $\sigma = 225$ K.}
    \label{fig:oxygen_ratio_star_planet}
\end{figure}

\autoref{fig:oxygen_ratio_star_planet}(a) illustrates the $(\text{O} - \text{Mg} - 2\ast\text{Si})/\text{Fe}$ ratio for planet(esimals) as a function of the stellar C/O ratio. In the Lodders03 system (C/O = 0.50), this ratio shows a subtle increase with higher $T_0$, indicating that oxygen availability increases at higher $T_0$. This trend reflects the oxygen-rich nature of solar-like systems, where Fe is predominantly oxidised, and the condensation sequence favours the incorporation of oxygen into silicates and oxides. In contrast, high C/O systems (HD94151 with C/O = 0.89 and HD24633 with C/O = 0.95) exhibit a reverse trend, with the ratio decreasing at higher $T_0$. The spread in values also correlates with the C/O ratio, with high C/O systems showing a broader range of $(\text{O} - \text{Mg} - 2\ast\text{Si})/\text{Fe}$ ratios, reflecting greater variability in their oxidation states. 

\autoref{fig:oxygen_ratio_star_planet}(b) showing the $\text{O}/(\text{N} + \text{C} + \text{Fe} + \text{Ni} + \text{S})$ ratio, further highlights the differences between these systems. In the Lodders03 system, oxygen remains relatively abundant, with the ratio exceeding 3 even at high $T_0$. This abundance ensures that Fe and other rock-forming elements are largely oxidized, resulting in smaller metallic cores. In high C/O systems, however, the oxygen availability decreases significantly with increasing C/O ratio, dropping below 2 at higher $T_0$. 
\onecolumn
\end{appendix}

\end{document}